%% file: main_Secure_trasceiver_design_MCC_V1.tex
\documentclass[12pt,draftclsnofoot,onecolumn]{IEEEtran}
\usepackage{setspace}
\IEEEoverridecommandlockouts
\usepackage{cite}
\usepackage{graphicx,graphics}
\graphicspath{{./figures/}}
\usepackage{subcaption}
\usepackage{amsmath,amssymb,amsfonts,amsthm}
\usepackage{algorithmic}
\usepackage[ruled,linesnumbered]{algorithm2e}
\usepackage{array}
\usepackage{url}
\usepackage[table,svgnames]{xcolor}

\newtheorem{proposition}{Proposition}

\newtheorem{lemma}{Lemma}

\newtheorem{assumption}{Assumption}

\title{Secure and Robust MIMO Transceiver for Multicast Mission Critical Communications%
\thanks{Supported by the EXTRANGE4G project with company ETELM funded by DGA. This work has been performed at LINCS laboratory.}}%
\author{\IEEEauthorblockN{Deepa Jagyasi and Marceau Coupechoux}\\
\IEEEauthorblockA{LTCI, Telecom Paris, Institut Polytechnique de Paris, France\\
 \{deepa.jagyasi,marceau.coupechoux\}@telecom-paris.fr}}%

\begin{document}

\maketitle

\begin{abstract}
\input{sections/abstract.tex}
\end{abstract}

\begin{IEEEkeywords}
mission critical communication,  physical layer security, robust transceiver design, imperfect CSI, norm-bounded error,  artificial-noise, MBSFN, SC-PTM, dynamic clustering.
\end{IEEEkeywords}

\section{Introduction}
\label{sec:introduction}
\input{sections/introduction.tex}
\section{Network and Transceiver Model}
\label{sec:system_model}
\input{sections/system_model.tex}

\section{Secure Transceiver design}
\label{sec:proposed_design}

\input{sections/proposed_design.tex}

\section{Numerical Results}
\label{sec:simulation_results}
\input{sections/simulation_results.tex}

\section{conclusion}
\label{sec:conclusion}
We propose a secure MIMO transceiver design for multi-BS multicast MCC that are resilient towards CSI errors following stochastic and norm-bounded error models. SMSE minimization problems are formulated under the constraint of maximum transmit power at every BS and minimum MSE at every eavesdropper. The BSs forming the coordinating cluster are obtained dynamically by using a greedy algorithm. Security is added in the system by optimal MIMO beamforming and by introducing an additional AN at the transmitters. The desired AN filter is jointly designed along with the precoder and receiver filters by solving the considered optimization problems using iterative and worst-case approaches. The performance is evaluated in terms of various parameters including security gap, BER and MSE. The computational analysis is also conducted and presented for both error models. Numerical results reveal the crucial role of robust designs for MCC, even in presence of norm-bounded errors. Adding AN degrades only slightly the performance of legitimate users but significantly improves the security of their communications. At last, we highlight the fact that increasing the number of cooperating BSs improves both reliability and security. However, dynamic clustering can represent a good trade-off if capacity becomes a requirement.  

\appendices

\section{Proof of Lemma~\ref{lem:ptmses}}
\label{app:prooflemma1}
\input{sections/appendix_lemma1}



\section{Proof for Propositions}
\label{app:alternativeW}
\input{sections/appendix_alternativeW.tex}




\bibliographystyle{IEEEtran}
\bibliography{ref.bib}

\end{document}

%% file: sections/abstract.tex
Mission-critical communications (MCC) involve all communications between people in charge of the safety of the civil society. MCC have unique requirements that include improved reliability, security and group communication support. In this paper, we propose a secure and robust Multiple-Input-Multiple-Output (MIMO) transceiver, designed for multiple Base Stations supporting multicast MCC in presence of multiple eavesdroppers. We formulate minimization problems with the Sum-Mean-Square-Error (SMSE) at legitimate users as an objective function, and a lower bound for the MSE at eavesdroppers as a constraint. Security is achieved thanks to physical layer security mechanisms, namely MIMO beamforming and Artificial Noise (AN). Reliability is achieved by designing a system which is robust to Channel State Information errors. They can be of known distribution or norm-bounded. In the former case, a coordinate descent algorithm is proposed to solve the problem. In the latter case, we propose an worst-case based iterative algorithm. Numerical results at physical layer and system level reveal the crucial role of robust designs for reliable MCC. We show the interest of both robust design and AN to improve the security gap. We show that full BS cooperation in preferred for highly secured and reliable MCC but dynamic clustering allows to trade-off security and reliability against capacity. 

%% file: sections/introduction.tex
\IEEEPARstart{M}
ission critical communications (MCC) are all communications between people in charge of the security and the safety of the civil society. Employees of public safety services, like policemen, firemen, rescue teams and ambulance nurses, but also from large companies managing critical infrastructures in the energy or transportation sectors require MCC for their operations~\cite{tcca}. MCC are conveyed by dedicated Private Mobile Radio (PMR) networks~\cite{MCC_services} that offer a group (or multicast) communication service. This is a one-to-many or many-to-many communication~\cite{3GPP_group_comm}, which is one of the most important features of PMR networks and is essential to manage teams of employees. Due to the critical aspects of their missions, MCC users also inherently require highly reliable and secure communications. In particular, sensitive information should not leak to unintended receivers although the broadcast nature of the wireless channel makes the network vulnerable to malicious eavesdroppers. In this context, we propose in this paper a physical layer secured Multiple-Input-Multiple-Output (MIMO) transceiver design for reliable multi-Base Stations (BS) multicast communication in the presence of malicious eavesdroppers.   

In the 3rd Generation Partnership Project (3GPP), group communication is based on Multimedia Broadcast/Multimedia Service (MBMS) standards~\cite{MCC_services}. It thus naturally benefits from the multicast transmission techniques~\cite{multicast_group_comm_lte}. In MBMS, the reliability is improved by coordinating multiple BSs within a so called synchronization area. When all BSs of the area cooperate, we have a Multimedia Multicast/Broadcast Single Frequency Network (MBSFN) transmission~\cite{MBSFN_SINR}. On the contrary, when BSs transmit independently, we have a Single-Cell Point-to-Multipoint (SC-PTM) transmission~\cite{3GPP}. Dynamic clustering, offering a good trade-off between MBSFN and SC-PTM is gaining popularity in the literature~\cite{dynamic_clus1_greedy,alaa_dynamic_clus_mcc}. This motivates our scenario of a multicast transmission from multiple BSs towards a group of users and provides us with a framework for system level evaluations. 

In order to ensure secure communications in presence of eavesdroppers, we rely on physical layer security~\cite{Wyner} mechanisms. They have the advantage of being independent of the secret key generation and distribution~\cite{PLS_to_cyrptography}. Although the use of long and complex keys is considered as one of the important techniques against eavesdroppers, the advent of powerful computational devices makes this approach indeed vulnerable on the long term~\cite{PLS_survey_5G}. In this paper, we consider physical layer security-based transceiver design for MCC by exploring signal processing methodologies in the presence of multiple eavesdroppers. Specifically, we incorporate security in two ways: MIMO beamforming is used to achieve the desired performance gain at legitimate users while degrading eavesdroppers channel; and artificial noise (AN) is added at the transmitter to guarantee additional security over the designed transceivers.

In our design, we formulate a problem in which the Sum-Mean-Square-Error (SMSE) should be minimized at legitimate users. This is an approach different from the traditional rate maximization problem, motivated by the reliability requirement in MCC. With the same goal, we propose a design that is robust to Channel State Information (CSI) errors. CSI is indeed never perfectly known due to various reasons such that estimation errors, feedback delays or pilot contamination. CSI errors thus affect the reliability of the communication~\cite{csi_effect}. Hence, it is crucial to design schemes that are resilient to such CSI imperfections. In this paper, we design systems that are robust to either stochastic errors or norm-bounded errors~\cite{Statictical_robust, convex_conic_NBE}. Stochastic Error (SE) models are defined based on the historical data observed in the system and assume a known Gaussian distribution. Norm-Bounded Errors (NBE) are considered to be bounded within an ellipsoid or spherical region without further information on the statistics of the errors~\cite{convex_conic_NBE}. We perform a comparative analysis of both models in terms of system performance.

\subsection{Related Work}

Physical layer security has been investigated for various communication applications, most of which assume a simple wiretap communication channel model~\cite{Wyner, Shannon}. In this setting, one legitimate transmitter (Alice) communicates with one legitimate receiver (Bob) (thus in unicast) in the presence of a single eavesdropper (Eve). 
Information theoretic aspects of secrecy have been widely studied in the literature, see e.g.~\cite{secrecy_rate_IT1,secrecy_rate_IT2}, our work however deals with signal processing techniques to achieve secure communications~\cite{signal_processing_security_book}. 
From the signal processing perspective, physical layer security has been studied for simple wiretap channels in various contexts such as artificial-noise aided security~\cite{AN_wiretap_non_linear,an_tx_correlation,an_miso_wiretap}, secure beamforming techniques~\cite{beamforming_wiretap, beamforming_relay}, or diversity oriented security~\cite{diversity_relay}. However, secured designs considering complex communication scenarios involving multiple transmitters, receivers and eavesdroppers have been observed in the literature only over the past decade. For example, uplink multiuser transmissions are considered in \cite{Alamouti_AN_bounds} and relay-assisted security is studied in~\cite{joint_relay_downlink_AN_cai}. 
The multicast scenario has been studied in~\cite{secure_multicast_perfectCSI,multicast_joint,multicast_multiple_eve,an_multigroup_multicast}, however always while assuming a single transmitting BS. To the best of our knowledge, secured multiple BS multicast system design has not been reported in the literature. Specifically in MCC, physical layer security has been considered in~\cite{MCC_allocation_urllc, MCC_secrecy_authentication1, MCC_secrecy_authentication2} in the context of resource allocation resource allocation~\cite{MCC_allocation_urllc} or for authentication~\cite{MCC_secrecy_authentication1,MCC_secrecy_authentication2}, but only for machine-type communications. It is however worthwhile noting that no instance addressing the design of secure transceivers for MCC group communications has been observed so far.  


MCC group communications involve text, image, audio or video exchanges in multicast. In this context, maximizing the data rate or the secrecy rate, as it is done usually in the literature, is not the main objective of operators. Instead, together with the security, the correctness of the data is of utmost importance. In our work, we thus consider a secure SMSE minimization-based transceiver design in the presence of multiple eavesdroppers. In the literature, only two instances of minimum MSE-based secure precoder design have been respectively discussed in~\cite{mmse_linear_wiretap} and \cite{mmse_non_linear}, however, for a simple wiretap communication scenario. These designs cannot be readily adapted for the proposed system due to increased complexity related to the presence of multiple coordinating BSs and eavesdroppers. Moreover, the multicast transmission, which consists of transmitting a common message to all legitimate users, makes the processing at eavesdroppers easier and thus requires a specific design. 

The addition of AN for the secure design of communication is either done in the null space of legitimate users, see e.g.~\cite{an_miso_wiretap,AN_wiretap_non_linear}, or is jointly designed with the precoder as in~\cite{joint_relay_downlink_AN_cai}. In this paper, we adopt a joint AN and transceiver design approach, where an AN shaping matrix is designed by solving the joint optimization problem and meet the overall design constraints. We confirm the interest of such a technique in the specific scenario of MCC. 


At last, to improve the system performance under realistic channel uncertainties, robustness needs to be incorporated as part of the design. Effectiveness of this approach is well-studied in the literature for various wireless communication applications~\cite{robust_relay2,robust_downlink,robust_relay_wireless}. However, most existing works on the design of physical-layer secured transceivers consider the availability of perfect CSI knowledge of both legitimate users and eavesdroppers. Very few papers are considering imperfect CSI~\cite{robust_wiretap_secure, robust_swipt_secure}, and only for the wiretap communication channel. In this paper, we consider imperfect CSI at both legitimate users and eavesdroppers in a complex scenario involving multiple BSs. A robust design is proposed for both stochastic error (when the error statistics can be learned) and norm-bounded error models (when minimal prior knowledge is available).   


A preliminary version of this paper has been published in~\cite{self_MCC_multiBS_multicast}. This reference neither include the security issue nor the norm-bounded error model. The system level performance evaluation is based on the work of~\cite{alaa_dynamic_clus_mcc}, but this reference, which proposes a clustering algorithm for MCC, does not implement any MIMO transceiver design.   

\subsection{Contribution}

In this paper, we propose a physical layer secured and  robust MIMO transceiver design for multi-BS multicast MCC system in presence of multiple eavesdroppers. The main contributions of this work are summarized as follows:

\begin{itemize}
    \item We formulate novel SMSE minimization problems to capture the reliability and security requirements of multicast MCC. Specifically, two optimization problems (\ref{P1} and \ref{P2}) are considered according to the type of CSI errors, i.e., stochastic (Assumption~\ref{ass:stochasticerror}) and norm-bounded errors (Assumption~\ref{ass:nbeerror}). Security aspects are tackled using MIMO beamforming and AN and accounted in the miminization problems as a lower bound constraint for the MSE of eavesdroppers. 
    \item When stochastic errors are assumed, we propose a coordinate descent iterative algorithm to solve the SMSE minimization problem (Algorithm~\ref{alg:iter_algo}). The algorithm is based on closed-form equations for the MSE (Lemma~\ref{lem:ptmses}) and the gradient of the Lagrangian (Proposition~\ref{prop:stochasticerr}). 
    \item When norm-bounded errors are assumed, we adopt a worst-case approach and decompose the original problem into three sub-problems. Resultant robust filters and AN shaping matrix are obtained by sequentially solving individual sub-problems in an iterative way (Algorithm~\ref{alg:iter_norm}).
    \item We provide numerical results at physical layer and system level to gain insights for the proposed designs. Physical layer simulations show the importance of robust designs for ensuring highly reliable MCC, even when norm-bounded errors are present. We show also the interest of AN for multicast MCC to ensure secure communications. System level simulations reveal that a full cooperation of the BSs of the synchronization area is preferred for reliable and secured MCC. If capacity becomes an important consideration, dynamic clustering can be adopted at the expense of less secured and reliable communications.  
\end{itemize}
The paper is structured as follows: Sec.~\ref{sec:system_model} describes the network and transceiver models. The problem formulations and the design of the transceivers are presented in Sec.~\ref{sec:proposed_design}. Physical layer and system level simulations are shown in Sec.~\ref{sec:simulation_results}. Sec.~\ref{sec:conclusion} concludes the paper.

\noindent {\em Notations}: We use bold-faced lowercase letters to denote column vectors and bold-faced uppercase letters to denote matrices. For any matrix ${\bf X}$, ${\rm tr}({\bf X})$, $\mathbb{E}\{{\bf X}\}$, ${\bf X}^{H}$, and ${\bf X}^{T}$ denote trace, expectation, conjugate transpose, and transpose operator, respectively. 

%% file: sections/system_model.tex

\begin{figure}[t]
\centering
\includegraphics[scale=0.35]{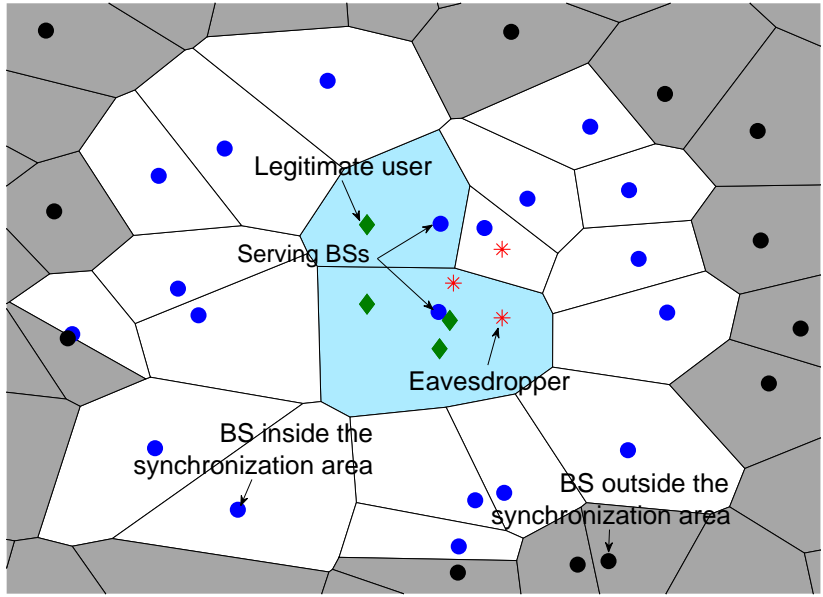}
\caption{Network model: White and blue cells form the MBSFN synchronization area; blue cells are serving a group of legitimate users (green diamonds) while a set of eavesdroppers (red crosses) overhear the multicast communication.}
\label{fig:net_mod}
\end{figure}
In this section, we present the network and transceiver models. 

\subsection{Network Model}
\label{subsec:network_model}
We consider the downlink of a cellular network dedicated to MCC with BSs serving legitimate users. Every legitimate user belongs to a multicast group, i.e., users of a given group receive the same information from the network. Every group is served by a cluster of coordinated BSs. In addition to legitimate users, we assume the potential presence of eavesdroppers, who listen to the transmitted information in a passive mode, i.e., without any tampering of the legitimate messages or active participation with the BSs. Fig.~\ref{fig:net_mod} shows such a cluster of BSs communicating with a set of group users. Among the BSs in the network, a set $\mathcal{B}$ of BSs are assumed to be synchronized: they operate at same frequency and utilize the same time/frequency resource block for communication. In the terminology of MBMS, $\mathcal{B}$ is called a {\it MBSFN synchronization area}. In Fig.~\ref{fig:net_mod}, white and blue cells form the MBSFN synchronization area, while grey cells are outside. 

In the proposed system, we consider a dynamic coordinated cluster of BSs $\mathcal{S}\subseteq \mathcal{B}$ (in blue in Fig.~\ref{fig:net_mod}), for every group of users $\mathcal{U}$. When $\mathcal{S} = \mathcal{B}$, all BSs of the MBSFN synchronization area cooperate to serve the group, this is called a {\it full MBSFN transmission}. Cells outside $\mathcal{B}$ contribute to the co-channel interference. When $|\mathcal{S}|=1$, we have a {\it SC-PTM transmission}. In general, a subset $\mathcal{S} \subseteq \mathcal{B}$ is dynamically selected for every group. In this case, cells in $\mathcal{S}$ cooperate, while cells in $\mathcal{B}\backslash\mathcal{S}$ and cells outside the MBSFN synchronization area contribute to the co-channel interference. In this paper, we consider a greedy clustering algorithm, where the cluster is formed by progressively selecting $K_T'$ best BS on the basis of maximum Signal to Interference plus Noise Ratio (SINR) achieved at the group users (see Algorithm~\ref{alg:greedy}). We first include in $\mathcal{S}$ the BSs that provide the highest receive power to every user. If $|\mathcal{S}|\leq K_T'$, we complete with $K_T'-|\mathcal{S}|$ BSs providing the highest sum SINR to group users. $K_T'$ is considered as a design parameter that controls the minimum cluster size. The greedy clustering algorithm and its variants are widely adopted in the literature related to multi-point cooperation, see e.g. \cite{dynamic_clus1_greedy,greedy_clustering_uplink}. We denote $K_T$ as the number of BSs eventually selected by Algorithm~\ref{alg:greedy}.


\begin{algorithm}[t]
\caption{Greedy Clustering}
\label{alg:greedy}
\begin{algorithmic}[1]
\STATE {\bf Input:} Locations of BSs and group users, ${K_T'}\leq |\mathcal{B}|$: minimum cluster size
\STATE {\bf Init:} $\mathcal{S}\gets \emptyset$
\FOR{every user}
    \STATE Find the BS $t$ providing the highest receive power
    \STATE $\mathcal{S}\gets \mathcal{S}\cup \{t\}$
\ENDFOR
\IF{$|\mathcal{S}|<{K_T'}$}
    \STATE Find the set $\mathcal{S}'$ of ${K_T'}-|\mathcal{S}|$ BSs maximizing the sum SINR for group users
    \STATE $\mathcal{S}\gets \mathcal{S}\cup \mathcal{S}'$
\ENDIF
\RETURN $\mathcal{S}$
\end{algorithmic}
\end{algorithm}

\begin{figure}
\centering
\includegraphics[scale=0.48]{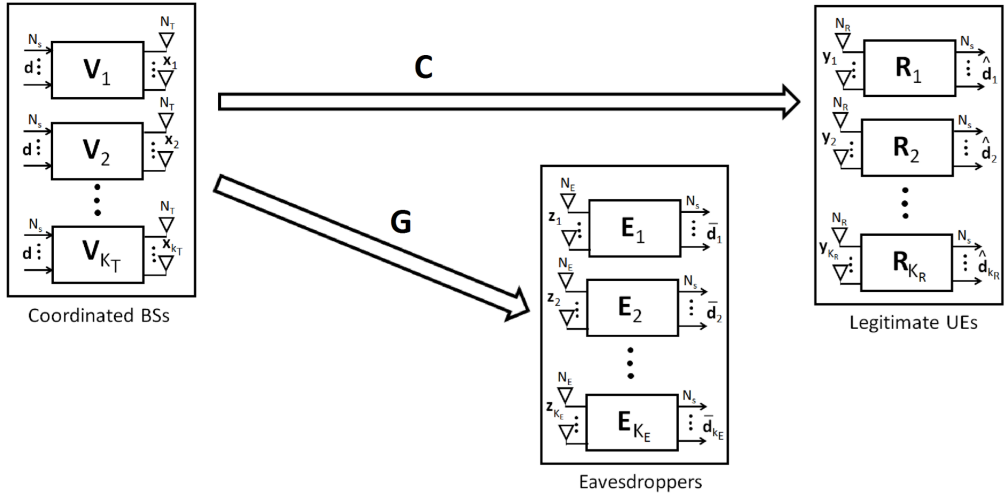}
\caption{System diagram for multi-cast scenario in mission critical communication in the presence of passive Eavesdroppers.}
\label{fig:sys_mod}
\end{figure}
\subsection{Transceiver Model}
\label{subsec:transceiver_model}
To analyze the transmission towards a group of users, we consider the secure multi-user MIMO multicast wireless communication scenario as shown in Fig.~\ref{fig:sys_mod}, where the $K_T$ BSs of cluster $\mathcal{S}$  multicast a common message to $K_R$ legitimate user-equipments (UEs) in a group. 
The transmitted signal is assumed to be overheard by $K_E$ passive eavesdroppers (eves). All the nodes in the system are considered to be equipped with MIMO processing, where each BS, UE and eve have $N_T$, $N_R$ and $N_E$ antennas respectively. Each BS multicasts a time-slotted $N_s$ dimensional column vector ${\bf d}$ with transmit power $P_T$, where $N_s$ is the number of parallel data streams transmitted by the BS. The data ${\bf d}$ is considered to be mutually independent, so that $\mathbb{E}[\mathbf{d}\mathbf{d}^H]=\mathbf{I}_{N_s}$. Before transmission, the data vector is processed by a $(N_T \times N_s)$ dimensional precoder matrix ${\bf V}_{t}$ at the $t$-th BS, $ t=1,\cdots,K_T$. In order to improve security, we introduce an additional artificial noise vector ${\bf z}_t$ of size $(N_T\times 1)$ with zero mean and variance $\mathbb{E}[{\bf z}_{t}{\bf z}_{t}^{H}]=\mathbf{\sigma}_{zt}^2{\bf I}_{N_T}$ at the $t$-th transmitter. The presence of AN has the goal of depleting the information leak to eavesdroppers. Furthermore, an AN-shaping matrix ${{\bf W}_t}$ of size $(N_T\times N_T)$ is considered to regulate the effect of AN in the overall design. Transceivers and AN-shaping matrix are jointly designed and this information is supposed to be shared between the transmitters and the legitimate users. 
Hence the signal transmitted from $t$-th BS is given by:
\begin{eqnarray}
{{\bf x}_t} = {{\bf V}_t}{\bf d} + {{\bf W}_t}{{\bf z}_t},
\end{eqnarray}
and the total transmit power at $t$-th BS is given by:
\begin{eqnarray}
\label{eq:transmit_power_tth_BS_basic}
{P}_{t} &\triangleq&\mathbb{E}[||{\bf x}_{t}{\bf x}_{t}^{H}||]. 
\end{eqnarray}
We denote the channel gain between the $t$-th BS and the $l$-th legitimate UE and between the $t$-th BS and the $e$-th eavesdropper by ${\bf C}_{tl}$ (with dimension $N_R \times N_T$) and ${\bf G}_{te}$ of dimension ($N_E \times N_T$), respectively. We assume quasi-static Rayleigh fading channels that remain static over one transmission time-slot. Consequently, the received signal ${\bf y}_{l}$ at legitimate UE $l$ is given by:
\begin{eqnarray}
{\bf y}_{l} = {\bf C}_{tl}{{\bf V}_{t}}{\bf d} + {\bf C}_{tl}{{\bf W}_{t}}{{\bf z}_t} + {\bf n}_{l},
\end{eqnarray}
where ${\bf n}_l$ is the $N_R$-dimensional zero mean random white Gaussian noise vector at the $l$-th UE's receive antennas with $\mathbb{E}[\mathbf{n}_l\mathbf{n}_l^H]=\sigma_{nl}^{2}{\bf I}_{N_R}$. The random noise vector is uncorrelated with the data vector, so that $\mathbb{E}[\mathbf{n}_l\mathbf{d}^H]=0$. The received signal at the UE $l$ is estimated as $\widehat{\bf d}_{l}$ (of dimension $N_s \times 1$) after passing through a $N_R \times N_s$ dimensional receive filter matrix ${\bf R}_{l}$. The estimated data is given by:
\begin{eqnarray}
\widehat{\bf d}_{l} = {\bf R}_{l}{\bf C}_{tl}{{\bf V}_{t}}{\bf d} + {\bf R}_{l}{\bf C}_{tl}{{\bf W}_{t}}{{\bf z}_t} + {\bf R}_{l}{\bf n}_{l}.
\end{eqnarray}
Thus, the MSE at the $l$-th legitimate UE is given by:
\begin{eqnarray}
\label{eq:MSE_lth_UE_basic}
{\epsilon}_{l} &\triangleq& \mathbb{E}[||{\bf d}-\widehat{\bf d}_{l}||^{2}].
\end{eqnarray}
Similarly at the eavesdroppers, the received signal ${\bf y}_{e}$ at the $e$-th eavesdropper can be given by:
\begin{eqnarray}
{\bf y}_{e} = {\bf G}_{te}{{\bf V}_{t}}{\bf d} + {\bf G}_{te}{{\bf W}_{t}}{{\bf z}_t} + {\bf n}_{e},
\end{eqnarray}
where ${\bf n}_e$ is the random white Gaussian noise vector of size $N_E \times 1$ at the $e$-th eavesdropper's antenna elements with zero mean and covariance $\mathbb{E}[\mathbf{n}_e\mathbf{n}_e^H]=\sigma_{ne}^{2}{\bf I}_{N_E}$. The random noise vector is uncorrelated with data vector such that $\mathbb{E}[\mathbf{n}_e\mathbf{d}^H]=0$. 
In this work, we assume that eavesdropper  implements a classical Minimum-Mean-Square-Error (MMSE) linear receive filter. However, it can replaced with any other linear receiver model. The considered MMSE receive filter at $e$-th eavesdropper is given as:
\begin{eqnarray}
{\bf E}_{e} &=& \Big(\sum_{t=1}^{K_T}{\bf V}_{t}^{H}{\bf G}_{te}^{H}\Big)\Big(\sum_{t=1}^{K_T}{\bf G}_{te}{\bf V}_{t}{\bf V}_{t}^{H}{\bf G}_{te}^{H} + {\sigma}_{ne}^{2}{\bf I} \Big)^{-1}. \label{eq:eve_mmse_filter}
\end{eqnarray}
Eavesdroppers do not have the information about the presence of AN in the received signal and hence ${\bf W}_{t}{\bf z}_{t}$ is not considered in the MMSE receive filter design. 
After passing ${\bf y}_e$ through the $N_E \times N_s$ receive filter ${\bf E}_{e}$, the estimated data $\overline{{\bf d}}_{e}$ at the $e$-th eavesdropper is given by:
\begin{eqnarray}
\overline{\bf d}_{e} = {\bf E}_{e}{\bf G}_{te}{{\bf V}_{t}}{\bf d} + {\bf E}_{e}{\bf G}_{te}{{\bf W}_{t}}{{\bf z}_{t}} + {\bf E}_{e}{\bf n}_{e}.
\end{eqnarray}
Thus, the MSE at the $e$-th eavesdropper can be obtained as:
\begin{eqnarray}
\label{eq:MSE_eth_eve_basic}
{\epsilon}_{e} &\triangleq& \mathbb{E}[||\bf{d}-\overline{\bf{d}}_{e}||^{2}].
\end{eqnarray}
Furthermore, we assume imperfect CSI knowledge at the receivers as follows.
The erroneous channel between the $t$-th BS and the $e$-th eve is modeled as:  
\begin{IEEEeqnarray}{lll}
{\bf G}_{te}={\widehat{\bf G}_{te}}+{\boldsymbol{\Delta}_{te}},
\end{IEEEeqnarray}
where $\widehat{\bf G}_{te}$ is the available estimate of CSI and $\boldsymbol{\Delta}_{te}$ refers to the corresponding channel uncertainties. 
In the same way, the CSI knowledge of legitimate users may not always be perfect, so that a robust transceiver designed is required.  
The erroneous channel between the $t$-th BS and the $l$-th user is modeled as:
\begin{IEEEeqnarray}{lll}
{\bf C}_{tl}={\widehat{\bf C}_{tl}}+{\boldsymbol{\Delta}_{tl}},
\end{IEEEeqnarray}
where $\widehat{\bf C}_{tl}$ is the estimated channel and $\boldsymbol{\Delta}_{tl}$ corresponds to channel uncertainties.
We consider two ways of modeling the error $\boldsymbol{\Delta}_{te}$ and $\boldsymbol{\Delta}_{tl}$. The first one assumes that the errors statistics have been learned from previous measurements. The second is valid when only a rough estimate of the noise power is available. 

\begin{assumption}[Stochastic error model] \label{ass:stochasticerror}
CSI errors $\boldsymbol{\Delta}_{te}$ and $\boldsymbol{\Delta}_{tl}$ are modeled as Gaussian random variables such that $\mathbb{E}[\boldsymbol{\Delta}_{te}\boldsymbol{\Delta}_{te}^{H}] = \sigma_{te}^{2}{\bf I}$ and $\mathbb{E}[\boldsymbol{\Delta}_{tl}\boldsymbol{\Delta}_{tl}^{H}] = \sigma_{tl}^{2}{\bf I}$.
\end{assumption}

\begin{assumption}[Norm-bounded error model] \label{ass:nbeerror}
CSI errors $\boldsymbol{\Delta}_{te}$ and $\boldsymbol{\Delta}_{tl}$ are modeled using the norm-bounded error (NBE) model, also known as deterministic-bounded error model~\cite{robust_relay_wireless}, where $\boldsymbol{\Delta}_{te}$ and $\boldsymbol{\Delta}_{tl}$ are respectively taken in continuous sets, called uncertainty regions, defined by:
\begin{eqnarray}
\mathcal{G}_{te} &=& \{{\bf \Delta}_{te} : {||{\bf \Delta}_{te}||}^{2} \leq {\tau}_{te} \}, \\
\mathcal{C}_{tl} &=& \{{\bf \Delta}_{tl} : {||{\bf \Delta}_{tl}||}^{2} \leq {\tau}_{tl} \},
\end{eqnarray}
where ${\tau}_{te}$ and ${\tau}_{tl}$ denote the radii of the uncertainty regions.
\end{assumption}
 
The channel errors for both legitimate UEs and eves are considered to be uncorrelated with the transmitted data sequence as well as to the additive white noise vector, i.e., $\mathbb{E}[{\bf d}\boldsymbol{\Delta}_{t,l}] = 0$, $\mathbb{E}[{\bf n}_{l}\boldsymbol{\Delta}_{t,l}] = 0$ for all $t,l$ and $\mathbb{E}[{\bf d}\boldsymbol{\Delta}_{t,e}] = 0$, $\mathbb{E}[{\bf n}_{e}\boldsymbol{\Delta}_{t,e}] = 0$ for all $t,e$. We now specify that the expectations in \eqref{eq:MSE_lth_UE_basic} and~\eqref{eq:MSE_eth_eve_basic} are considered over data, channel matrix, noise and estimation errors.


%% file: sections/proposed_design.tex
In this section, we present our robust and secure transceiver design. 


\subsection{Stochastic CSI Errors} \label{subsec:stocherrors}

Our goal is to obtain the optimal precoder, receive filter, and AN-shaping matrices ${\bf V}_{t}$, ${\bf R}_{l}$, and ${\bf W}_{t}$ for secure communications at all the BSs and legitimate UEs while minimizing the overall SMSE of the legitimate UEs under the constraint of a maximum transmit power at every BS and a minimum MSE for every eavesdropper. In this sub-section, Assumption~\ref{ass:stochasticerror} is considered. 
Our joint optimization problem can thus be formulated as follows: 
\begin{equation}\tag{$\mathcal{P}_1$}\label{P1}
\begin{aligned}
& \underset{\substack{\mathbf{V}_{t},\mathbf{W}_{t},\mathbf{R}_{l}\\ t=1...K_T, l=1...K_R}}{\text{minimize}}
& & \sum_{l=1}^{K_R}{\epsilon}_{l}, \\
& \text{subject to}
& & C1:~{\epsilon}_{e} \geq \Gamma, \qquad \forall e \in \{1,\cdots,K_{E}\}, \\
&&& C2:~P_{t}\leq P_T, \qquad \forall t \in \{1,\cdots,K_{T}\}, \\
\end{aligned}
\end{equation}
The MSE $\epsilon_l$ and $\epsilon_e$ at the legitimate user $l$ and eavesdropper $e$ are obtained using \eqref{eq:MSE_lth_UE_basic} and \eqref{eq:MSE_eth_eve_basic}, respectively. The transmit power $P_t$ at $t$-th BS is given by \eqref{eq:transmit_power_tth_BS_basic}. In $C1$, $\Gamma$ is a design parameter that represents the lower bound on the achievable MSE expected at each eavesdropper. In $C2$, $P_T$ is the maximum transmit power at every BS. 

\begin{lemma} \label{lem:ptmses}
With Assumption~\ref{ass:stochasticerror}, we have the following result:
\begin{eqnarray}
P_t&=& {\rm}{tr}({\bf V}_{t}{\bf V}_{t}^{H}+\sigma_{zt}^{2}{\bf W}_{t}{\bf W}_{t}^{H}), \\
{\epsilon}_{l} &=& {\rm tr}({\bf I}) -%
{\rm tr}(\sum_{t=1}^{K_T}{\bf R}_{l}\widehat{\bf C}_{tl}{\bf V}_{t}) -%
{\rm tr}(\sum_{t=1}^{K_T}{\bf V}_{t}^{H}\widehat{\bf C}_{tl}^{H}{\bf R}_{l}^{H}) +%
{\rm tr}(\sum_{t=1}^{K_T}{\bf R}_{l}\widehat{\bf C}_{tl}{\bf V}_{t}{\bf V}_{t}^{H}\widehat{\bf C}_{tl}^{H}{\bf R}_{l}^{H})  \nonumber \\%
&+& {\rm tr}(\sum_{t=1}^{K_T}\sigma_{zt}^2{\bf R}_{l}\widehat{\bf C}_{tl}{\bf W}_{t}{\bf W}_{t}^{H}\widehat{\bf C}_{tl}^{H}{\bf R}_{l}^{H}) +%
{\sigma}_{nl}^{2}{\rm tr}({\bf R}_{l}{\bf R}_{l}^{H}) +%
\sum_{t=1}^{K_T}{\sigma}_{tl}^{2}{\rm tr}({\bf R}_{l}{\bf R}_{l}^{H}){\rm tr}({\bf V}_{t}{\bf V}_{t}^{H}) \nonumber \\%
&+&  \sum_{t=1}^{K_T}{\sigma}_{tl}^{2}\sigma_{zt}^2{\rm tr}({\bf R}_{l}{\bf R}_{l}^{H}){\rm tr}({\bf W}_{t}{\bf W}_{t}^{H}), \label{eq:mse_leg} \\
{\epsilon}_{e}&=& {\rm tr}({\bf I}) -%
{\rm tr}(\sum_{t=1}^{K_T}{\bf E}_{e}\widehat{\bf G}_{te}{\bf V}_{t}) -%
{\rm tr}(\sum_{t=1}^{K_T}{\bf V}_{t}^{H}\widehat{\bf G}_{te}^{H}{\bf E}_{e}^{H}) +%
{\rm tr}(\sum_{t=1}^{K_T}{\bf E}_{e}\widehat{\bf G}_{te}{\bf V}_{t}{\bf V}_{t}^{H}\widehat{\bf G}_{te}^{H}{\bf E}_{e}^{H})  \nonumber \\%
&+&\quad {\rm tr}(\sum_{t=1}^{K_T}\sigma_{zt}^2{\bf E}_{e}\widehat{\bf G}_{te}{\bf W}_{t}{\bf W}_{t}^{H}\widehat{\bf G}_{te}^{H}{\bf E}_{e}^{H}) +%
{\sigma}_{ne}^{2}{\rm tr}({\bf E}_{e}{\bf E}_{e}^{H}) +%
\sum_{t=1}^{K_T}{\sigma}_{te}^{2}{\rm tr}({\bf E}_{e}{\bf E}_{e}^{H}){\rm tr}({\bf V}_{t}{\bf V}_{t}^{H}) \nonumber \\%
&+&  \sum_{t=1}^{K_T}{\sigma}_{te}^{2}\sigma_{zt}^2{\rm tr}({\bf E}_{e}{\bf E}_{e}^{H}){\rm tr}({\bf W}_{t}{\bf W}_{t}^{H}). \label{eq:mse_eve}
\end{eqnarray}
\end{lemma}

\begin{IEEEproof}
See Appendix~\ref{app:prooflemma1}.
\end{IEEEproof}

\begin{proposition} \label{prop:stochasticerr}
With Assumption~\ref{ass:stochasticerror},
the optimal transceiver and AN shaping matrices verify:
\begin{eqnarray}
{\bf V}_{t} &=& ({\bf A}_{t})^{-1}\Big(\sum_{l=1}^{K_R}\widehat{\bf C}_{tl}^{H}{\bf R}_{l}^{H} - \sum_{e=1}^{K_E}\lambda_{e}\widehat{\bf G}_{te}^{H}{\bf E}_{e}^{H}\Big) \label{eq:pre_robust}, \\
{\bf W}_t&=&{\bf B}_{t}/\sqrt{\mathbb[\rm tr({\bf B}_{t}{\bf B}_{t}^{H})]} \label{eq:an_robust}, \\
{\bf R}_{l} &=& \Big(\sum_{t=1}^{K_T}{\bf V}_{t}^{H}\widehat{{\bf C}}_{tl}^{H}\Big)\Big(\sum_{t=1}^{K_T}\widehat{{\bf C}}_{tl}{\bf V}_{t}{\bf V}_{t}^{H}\widehat{{\bf C}}_{tl}^{H} +%
\sum_{t=1}^{K_T}\sigma_{zt}^{2}\widehat{{\bf C}}_{tl}{\bf W}_{t}{\bf W}_{t}^{H}\widehat{{\bf C}}_{tl}^{H} + {\sigma}_{nl}^{2}{\bf I} \nonumber \\ 
&+&\sum_{t=1}^{K_T}{\sigma}_{tl}^{2}{\rm tr}({\bf V}_{t}{\bf V}_{t}^{H}){\bf I} +%
\sum_{t=1}^{K_T}{\sigma}_{tl}^{2}{\sigma}_{zt}^{2}{\rm tr}({\bf W}_{t}{\bf W}_{t}^{H}){\bf I} \Big)^{-1} \label{eq:rec_robust},
\end{eqnarray}
where 
\begin{eqnarray}
{\bf B}_{t}\!\!&=&\!\!{\bf I}-{\bf A}_{t}^{H}({\bf A}_{t}{\bf A}_{t}^{H})^{-1}{\bf A}_{t}, \\
{\bf A}_{t}\!\!\!\!&=&\!\!\!\! \sum_{l=1}^{K_R}\widehat{{\bf C}}_{tl}^{H}{\bf R}_{l}^{H}{\bf R}_{l}\widehat{\bf C}_{tl}\!\! +\!\!\! \sum_{l=1}^{K_R}{\sigma}_{tl}^{2}{\rm tr}({\bf R}_{l}{\bf R}_{l}^{H})\!\!-\!\!\sum_{e=1}^{K_E}{\lambda}_{e}\widehat{\bf G}_{te}^{H}{\bf E}_{e}^{H}{\bf E}_{e}\widehat{\bf G}_{te}\!\!  -\!\!\!\sum_{e=1}^{K_E}{\lambda}_{e}{\sigma}_{te}^{2}{\rm tr}({\bf E}_{e}{\bf E}_{e}^{H})\!\!+\!\! \lambda_{t}^{'}{\bf I}
\end{eqnarray}
and where $\lambda_{e} \geq 0$, and $\lambda_{t}^{'} \geq 0$ are the Lagrangian multipliers, which are calculated such that the constraints $C1$, and $C2$ are satisfied respectively.
\end{proposition}
\begin{IEEEproof}
See Appendix~\ref{app:alternativeW}.
\end{IEEEproof}
\begin{algorithm}[t]
\caption{Iterative procedure to obtain transceiver filters for stochastic errors}
\label{alg:iter_algo}
\begin{algorithmic}[1]
\STATE {\bf Input:} $\beta$, $K_T$, $K_R$, $K_E$, $\widehat{\mathbf{C}}_{tl}$, $\widehat{\mathbf{G}}_{te}$, $\sigma_{nl}$, $\sigma_{ne}$, $P_T$, $\Gamma$ $\forall$ $t\in\{1,\cdots,K_T\}$, $l\in\{1,\cdots,K_R\}$, and $e\in\{1,\cdots,K_E\}$
\STATE {\bf Init:} Randomly generate ${\bf V}_{t}$, ${\bf W}_{t}$ $\forall t\in\{1,\cdots,K_T\}$, $\epsilon_l' \gets 0$, $\epsilon_l\gets 0$, $\forall l\in\{1,\cdots,K_R\}$
\REPEAT
\STATE $\epsilon_l' \gets \epsilon_l$, $\forall l\in\{1,\cdots,K_R\}$
\STATE Update ${\bf E}_e$ $\forall e\in\{1,\cdots,K_E\}$ using \eqref{eq:eve_mmse_filter}
\STATE Update ${\bf R}_l$ using ${\bf V}_{t}$, ${\bf W}_{t}$ in \eqref{eq:rec_robust} $\forall l\in\{1,\cdots,K_R\}$
\STATE Solve for $\lambda_e$ and $\lambda_{t}^{'}$ using $C1$, $C2$  $\forall t\in\{1,\cdots,K_T\}$, and $\forall e\in\{1,\cdots,K_E\}$ 
\STATE Update ${\bf V}_t$ using $\lambda_e$, $\lambda_t^{'}$, ${\bf R}_{l}$, ${\bf E}_{e}$ in \eqref{eq:pre_robust} $\forall t=\{1,\cdots,K_T\}$ 
\STATE Update ${\bf W}_t$ using ${\bf V}_{t}$ in \eqref{eq:an_robust} $\forall t=\{1,\cdots,K_T\}$ 
\STATE Compute $\epsilon_l$ using ${\bf V}_{t}$, $\lambda_e$, $\lambda_t^{'}$,  ${\bf W}_t$, and ${\bf R}_{l}$ in \eqref{eq:mse_leg}
\UNTIL {$ |\epsilon_l - \epsilon_l'| \leq \beta $, $\forall l\in\{1,\cdots,K_R\}$ }
\RETURN ${\bf V}_{t}$, ${\bf W}_{t}$, $\forall t\in\{1,\cdots, K_T\}$, ${\bf R}_{l}$, $\epsilon_l$ $\forall l\in\{1,\cdots,K_R\}$
\end{algorithmic}
\end{algorithm}
From the proposition, we observe that even if the optimal problem is jointly non-convex, it is convex for each variable and an optimal solution can be obtained by finding an optimal solution for one variable while keeping others as fixed\cite{optimization_methods}. To obtain the solution, we utilize the coordinate descent method~\cite{coordinate_descent}, in which each optimization variable is optimized while considering other variables as fixed. The resulting iterative procedure is shown in Algorithm~\ref{alg:iter_algo}. In step 7, the Lagrange multipliers $\lambda_{e}, \forall e$ and $\lambda_{l}, \forall l$ are obtained by solving the set of non-linear equations from constraints $C1$ and $C2$ by using the function \texttt{fsolve} from Matlab, which implements a dogleg trust-region algorithm~\cite{optimization_methods}. Coordinate descent is known to generate sequences whose limit points are stationary if the objective function is continuously differentiable and the minimum along every coordinate is uniquely attained~\cite{bertsekas1997nonlinear}.

{\it Complexity analysis – } Let denote $N=\max(N_T, N_R, N_E, N_s)$ and $K=\max(K_T,K_R,K_E)$. $N$ is an upper bound on the number of antennas per device and the number of data streams; $K$ is related to the number of devices involved in the communication. The computation of ${\bf E}_e$ in \eqref{eq:eve_mmse_filter}, ${\bf R}_l$ in \eqref{eq:rec_robust}, ${\bf V}_t$ in \eqref{eq:pre_robust}, and ${\bf W}_t$ in \eqref{eq:an_robust} are dominated by $K$ inversions of matrices of size $N$, so that there complexity is in $O(KN^3)$. The computation of $\epsilon_l$ involves a sum of matrix multiplications and is thus in $O(KN^3)$. The dodleg algorithm implemented in \texttt{fsolve} is an iterative algorithm which achieves a $\varepsilon$-approximation of the solution with complexity $O(K^3\varepsilon^{-3})$ when using fully quadratic models in the derivative-free case with $2K$ unknowns~\cite{gratton2018complexity}.  
\begin{lemma}
Let $I_2$ be the number of iterations of Algorithm~\ref{alg:iter_algo} and $\varepsilon>0$ the accuracy of $\texttt{fsolve}$. The complexity of Algorithm~\ref{alg:iter_algo} is at most $O(I_2KN^3+I_2K^3\varepsilon^{-3})$.
\end{lemma}
In our simulations, $I_2=10$ iterations are generally sufficient to achieve convergence. 





\subsection{Norm-bounded CSI errors}
\label{sec:norm_bound}
In this sub-section, we consider Assumption~\ref{ass:nbeerror} for system design, i.e., we assume that channel state information errors are only norm bounded. In this case, the problem formulation can be written as:

\begin{equation}\tag{$\mathcal{P}_2$}\label{P2}
\begin{aligned}
& \underset{\substack{\mathbf{V}_{t},\mathbf{W}_{t},\mathbf{R}_{l}\\ t=1...K_T, l=1...K_R}}{\text{minimize}}
& & \sum_{l=1}^{K_R}{\epsilon}_{l}, \\
& \text{subject to}
& & C1, C2 \\
&&& C3:~\boldsymbol{\Delta}_{te} \in \mathcal{G}_{te}, \qquad \forall t \in \{1,\cdots,K_{T}\}, \forall e \in \{1,\cdots,K_{E}\}  \\ 
&&& C4:~\boldsymbol{\Delta}_{tl} \in \mathcal{C}_{tl}, \qquad \forall t \in \{1,\cdots,K_{T}\}, \forall l \in \{1,\cdots,K_{R}\} \\ 
\end{aligned}
\end{equation}

%
Note that $C3$ and $C4$ can be seen as an infinite number of constraints. In order to tackle this problem, we follow a worst-case approach, similar to the one presented in~\cite{shahbazpanahi2003robust}. In this approach, the SMSE of legitimate users is minimized for the worst-case error $\boldsymbol{\Delta}_{tl}$ subject to the constraint and, with the worst-case error $\boldsymbol{\Delta}_{te}$, the MSE of the eavesdroppers is maintained above the predefined threshold. This leads to this new formulation:
\begin{equation}\tag{$\bar{\mathcal{P}}_2$}\label{P2b}
\begin{aligned}
& \underset{\substack{\mathbf{V}_{t},\mathbf{W}_{t},\mathbf{R}_{l}\\ t=1...K_T, l=1...K_R}}{\text{minimize}}
& & \underset{\boldsymbol{\Delta}_{tl}\in \mathcal{C}_{tl}}{\max} \sum_{l=1}^{K_R}{\epsilon}_{l}, \\
& \text{subject to}
&& C2,\\
&&& C5:~\underset{\boldsymbol{\Delta}_{te} \in \mathcal{G}_{te}}{\min} \epsilon_e \geq \Gamma, \qquad \forall e \in \{1,\cdots,K_{E}\}
\end{aligned}
\end{equation}



In order to obtain a solution, we consider three sub-problems that are easier to solve individually. In every sub-problem, we assume that either the optimal transceiver matrices, or the worst-case errors $\boldsymbol{\Delta}_{tl}$ or $\boldsymbol{\Delta}_{te}$ are known and we optimize the other variables. This leads to an iterative algorithm, where, at every iteration, the three sub-problems are sequentially solved assuming the results of the previous iteration for the missing variables. 


\subsubsection{Sub-problem $\bar{\mathcal{P}}_2'$}
In the first sub-problem, we compute the optimal precoder ${\bf V}_{t}$, receive filter ${\bf R}_{l}$ and AN covariance matrix ${\bf W}_{t}$ while the worst-case channel errors ${\bf \Delta}_{te}$ and ${\bf \Delta}_{tl}$ are supposed to be known. The first optimization sub-problem can be thus written as:
\begin{equation}\tag{$\bar{\mathcal{P}}_2'$}\label{P2b1}
\begin{aligned}
& \underset{\substack{\mathbf{V}_{t},\mathbf{W}_{t},\mathbf{R}_{l}\\ t=1...K_T, l=1...K_R}}{\text{minimize}}
& & \sum_{l=1}^{K_R}{\epsilon}_{l}, \\
& \text{subject to}
&& C1, C2
\end{aligned}
\end{equation}
The optimization problem is similar to~\eqref{P1} and can be solved using the proof given in Appendix~\ref{app:alternativeW}  by replacing $\sigma_{tl}^{2}$ by $||{\bf \Delta}_{tl}||^{2}$, and $\sigma_{te}^{2}$ by $||{\bf \Delta}_{te}||^{2}$.


\subsubsection{Sub-problem $\bar{\mathcal{P}}_2''$}
In this sub-problem, the transceiver matrices and the worst-case error $\boldsymbol{\Delta}_{te}$ are supposed to be known and we look for the worst-case error $\boldsymbol{\Delta}_{tl}$. We can thus formulate the second sub-problem as:
\begin{equation}\tag{$\bar{\mathcal{P}}_2''$}\label{P2b2}
\begin{aligned}
& \underset{\substack{{\bf \Delta}_{tl}\\ t=1...K_T, l=1...K_R}}{\text{minimize}}
& & -\sum_{l=1}^{K_R}{\epsilon}_{l}, \\
& \text{subject to}
&& C4: {||{\bf \Delta}_{tl}||}^{2} \leq \tau_{tl} \qquad \forall t \in \{1,\cdots,K_{T}\}, \qquad \forall l \in \{1,\cdots,K_{R}\}
\end{aligned}
\end{equation}
The Lagrangian is given by:
\begin{eqnarray}
\label{eq:problem_b:lagrange}
{\rm L}({\bf \Delta}_{tl},{\kappa}_{tl}) &=& -\sum_{l=1}^{K_R} \epsilon_{l} + \sum_{t=1}^{K_T}\sum_{l=1}^{K_R}(\kappa_{tl}({||{\bf \Delta}_{tl}||}^{2} - \tau_{tl})), \nonumber \\
&=& -\sum_{l=1}^{K_R} \epsilon_{l} + \sum_{t=1}^{K_T}\sum_{l=1}^{K_R}(\kappa_{tl}({\rm tr}({\bf \Delta}_{tl}{\bf \Delta}_{tl}^{H}) - \tau_{tl})),
\end{eqnarray}
where $\kappa_{tl}\geq 0$ are the Lagrange multipliers associated to constraints $C4$. Considering~\eqref{eq:mse_leg}, it is observed that solving~\eqref{P2b2} is difficult. To simplify the solution, we consider an approximation of $\epsilon_{l}$ by ignoring the second and higher order terms of ${\bf \Delta}_{tl}$. Equating to zero, the partial derivative of the Lagrangian with respect to ${\bf \Delta}_{tl}^{H}$ and to $\kappa_{tl}$, respectively, we obtain:
\begin{eqnarray}
{\bf \Delta}_{tl} &=& \frac{1}{\kappa_{tl}}({\bf R}_{l}^{H}{\bf V}_{t}^{H} + {\bf R}_{l}^{H}{\bf R}_{l}\widehat{\bf C}_{tl}{\bf V}_{t}{\bf V}_{t}^{H} + \sigma_{zt}^{2}{\bf R}_{l}^{H}{\bf R}_{l}\widehat{\bf C}_{tl}{\bf W}_{t}{\bf W}_{t}^{H}),
\label{eq:problem_b:delta_sol} \\
\tau_{tl} &=& {\rm tr}({\bf \Delta}_{tl}{\bf \Delta}_{tl}^{H}). \label{eq:problem_b:tau}
\end{eqnarray}
Injecting \eqref{eq:problem_b:delta_sol} in \eqref{eq:problem_b:tau}, we obtain the Lagrange multipliers:
\begin{eqnarray}
\kappa_{tl} &=& \frac{1}{\sqrt{\tau_{tl}}}{||({\bf R}_{l}^{H}{\bf V}_{t}^{H} + {\bf R}_{l}^{H}{\bf R}_{l}\widehat{\bf C}_{tl}{\bf V}_{t}{\bf V}_{t}^{H} + \sigma_{zt}^{2}{\bf R}_{l}^{H}{\bf R}_{l}\widehat{\bf C}_{tl}{\bf W}_{t}{\bf W}_{t}^{H})||} \nonumber
\end{eqnarray}
Using the value of $\kappa_{tl}$ in~\eqref{eq:problem_b:delta_sol}, we get the following lemma.
\begin{lemma}
The worst-case error ${\bf \Delta}_{tl}$ that provides the maximum SMSE at the legitimate UEs for given optimal transceivers and AN while satisfying the norm-bound constraint $\tau_{tl}$ is given by:
\begin{eqnarray}
\label{eq:problem_b:delta_final}
{\bf \Delta}_{tl} &=& {\sqrt{\tau_{tl}}}\frac{({\bf R}_{l}^{H}{\bf V}_{t}^{H} + {\bf R}_{l}^{H}{\bf R}_{l}\widehat{\bf C}_{tl}{\bf V}_{t}{\bf V}_{t}^{H} + \sigma_{zt}^{2}{\bf R}_{l}^{H}{\bf R}_{l}\widehat{\bf C}_{tl}{\bf W}_{t}{\bf W}_{t}^{H})}{||({\bf R}_{l}^{H}{\bf V}_{t}^{H} + {\bf R}_{l}^{H}{\bf R}_{l}\widehat{\bf C}_{tl}{\bf V}_{t}{\bf V}_{t}^{H} + \sigma_{zt}^{2}{\bf R}_{l}^{H}{\bf R}_{l}\widehat{\bf C}_{tl}{\bf W}_{t}{\bf W}_{t}^{H})||}
\end{eqnarray}
\end{lemma}

\subsubsection{Sub-problem $\bar{\mathcal{P}}_2'''$}
\label{sec:norm_bound:eves}
In this third sub-problem, the transceiver matrices and the worst-case error $\boldsymbol{\Delta}_{tl}$ are supposed to be known and we look for the worst-case error $\boldsymbol{\Delta}_{te}$. It can be found by solving the problem defined in constraint $C5$. The sub-problem can thus be written as:

\begin{algorithm}[t]
\caption{Iterative procedure to obtain transceiver filters for norm-bounded errors}
\label{alg:iter_norm}
\begin{algorithmic}[1]
\STATE {\bf Input:} $\beta$, $K_T$, $K_R$, $K_E$, ${\bf E}_{e}$, ${\widehat{\mathbf{C}}_{tl}}$, ${\widehat{\mathbf{G}}_{te}}$, $\tau_{tl}$, $\tau_{te}$, $P_T$, $\Gamma$, $\forall$  $t\in\{1,\cdots,K_T\}$, $l\in\{1,\cdots,K_R\}$, and $e\in\{1,\cdots,K_E\}$ 
\STATE {\bf Init:} Randomly generate ${\bf V}_{t}$, ${\bf W}_{t}$, $\forall t\in\{1,\cdots,K_T\}$, ${\bf \Delta}_{tl} \in \mathcal{C}_{tl}$, ${\bf \Delta}_{te} \in \mathcal{G}_{te}$, $\epsilon_l\gets 0$, $\forall t\in\{1,\cdots,K_T\}$, $ l\in\{1,\cdots,K_R\}$, $ e\in\{1,\cdots,K_E\}$,  
\REPEAT
\STATE $\epsilon_l'\gets \epsilon_l$, $\forall l\in\{1,\cdots,K_R\}$
\STATE Solve $\bar{\mathcal{P}}_2'$ and update ${\bf V}_{t}$, ${\bf W}_{t}$, ${\bf R}_l$ using ${\bf \Delta}_{tl}$, ${\bf \Delta}_{te}$, ${\bf E}_{e}$ and Algorithm~\ref{alg:iter_algo} $\forall t\in\{1,\cdots,K_T\}$, $ l\in\{1,\cdots,K_R\}$, $e\in\{1,\cdots,K_E\}$ 
\STATE Solve $\bar{\mathcal{P}}_2''$ and update ${\bf \Delta}_{tl}$ using ${\bf V}_{t}$, ${\bf W}_{t}$, ${\bf R}_l$, in~\eqref{eq:problem_b:delta_final} $\forall t\in\{1,\cdots,K_T\}$, $ l\in\{1,\cdots,K_R\}$
\STATE Solve $\bar{\mathcal{P}}_2'''$ and update ${\bf \Delta}_{te}$ using ${\bf V}_{t}$, ${\bf W}_{t}$, ${\bf E}_{e}$, in~\eqref{eq:problem_c:delta_final} $\forall t\in\{1,\cdots,K_T\}$, $ e\in\{1,\cdots,K_E\}$
\STATE Compute $\epsilon_l$ using ${\bf V}_{t}$, ${\bf W}_t$, and ${\bf R}_{l}$, ${\bf \Delta}_{tl}$, ${\bf \Delta}_{te}$ in \eqref{eq:mse_leg}
\UNTIL {$| \epsilon_l - \epsilon_l'| \leq \beta $, $\forall l\in\{1,\cdots,K_R\}$}
\end{algorithmic}
\end{algorithm}
\begin{equation}\tag{$\bar{\mathcal{P}}_2'''$}\label{P2b3}
\begin{aligned}
& \underset{{\bf \Delta}_{te}, t=1...K_T}{\text{minimize}}
& & \epsilon_{e} \geq \Gamma, \\
& \text{subject to}
&& C3: {||{\bf \Delta}_{te}||}^{2} \leq \tau_{te} \qquad \forall t \in \{1,\cdots,K_{T}\}
\end{aligned}
\end{equation}
This problem is similar to~\eqref{P2b2}, the solution can thus be derived along the same lines. 
\begin{lemma}
The worst-case error ${\bf \Delta}_{te}$ that provides the minimum MSE at the eavesdroppers for given optimal transceivers and AN while satisfying the norm-bound $\tau_{te}$ is given by:
\begin{eqnarray}
\label{eq:problem_c:delta_final}
{\bf \Delta}_{te} = -\sqrt{\tau_{te}}\frac{({\bf E}_{e}^{H}{\bf V}_{t}^{H} + {\bf E}_{e}^{H}{\bf E}_{e}\widehat{\bf G}_{te}{\bf V}_{t}{\bf V}_{t}^{H} +\sigma_{zt}^{2} {\bf E}_{e}^{H}{\bf E}_{e}\widehat{\bf G}_{te}{\bf W}_{t}{\bf W}_{t}^{H})}{||({\bf E}_{e}^{H}{\bf V}_{t}^{H} + {\bf E}_{e}^{H}{\bf E}_{e}\widehat{\bf G}_{te}{\bf V}_{t}{\bf V}_{t}^{H} + \sigma_{zt}^{2}{\bf E}_{e}^{H}{\bf E}_{e}\widehat{\bf G}_{te}{\bf W}_{t}{\bf W}_{t}^{H})||}
\end{eqnarray}
\end{lemma}

A stationary solution for~\eqref{P2b} is now obtained by a three step iterative process as given in Algorithm~\ref{alg:iter_norm}. We first obtain the optimal solution considering that the channel errors are known. Afterwards, the worst case channel errors are computed considering the optimal solution is known. 

{\it Complexity analysis – } The computation of $\epsilon_l$ using \eqref{eq:mse_leg} is again in $O(KN^3)$. The computation of ${\bf \Delta}_{tl}$ and ${\bf \Delta}_{te}$ involves $K$ matrix multiplications and is thus $O(KN^3)$. The complexity of Algorithm~\ref{alg:iter_norm} is thus dominated by the inner loop constituted by Algorithm~\ref{alg:iter_algo}.

\begin{lemma}
Let $I_3$ be the number of iterations of Algorithm~\ref{alg:iter_norm}. The complexity of Algorithm~\ref{alg:iter_norm} is at most $O(I_3I_2KN^3+I_3I_2K^3\varepsilon^{-3})$.
\end{lemma}
In our simulations, we observe that generally $I_3=8$ iterations are sufficient to achieve convergence of the algorithm.

%% file: sections/simulation_results.tex
In this section, we illustrate the performance of our designs with numerical simulations at physical layer and at system level. 

\subsection{Physical Layer Simulations}
\label{sec:phy_simulations}
In physical layer simulations, there is no co-channel interference and both legitimate UEs and eavesdroppers experience the same path-loss. In results, we refer to the Non-Robust (NR) design, Robust (R)  design, Stochastic Errors (SE) and Norm-Bounded Errors (NBE). Unless otherwise specified, the simulation parameters are the following: $K_T=4$, $K_R=8$, $K_E=2$, $N_T=16$, $N_R=8$, $N_E=4$, and $N_s=2$; $P_T=0$~dBm; $\sigma_{nl}^2={P_T}/{\rm SNR}$, $l={1,2,..,K_R}$, $\sigma_{ne}^2={P_T}/{\rm SNR}$, $e={1,..,K_E}$, where ${\rm SNR}$ is the transmit ${\rm SNR}$. For the R-SE design, $\sigma_{tl}^2=0.04$ and $\sigma_{te}^2=0.09$. For the R-NBE design, $\tau_{tl}=0.04$ and $\tau_{te}=0.09$. AN variance is $\sigma_{zt}^2=0.09$. The target MSE threshold at each eavesdropper is $\Gamma=0.5$. We assume a QPSK modulation and average performance metrics over $10^6$ data samples for every simulation. In Algorithm~\ref{alg:iter_algo} and~\ref{alg:iter_norm}, $\beta=10^{-4}$. 

\begin{figure}[t]
\centering
\includegraphics[width=0.6\textwidth]{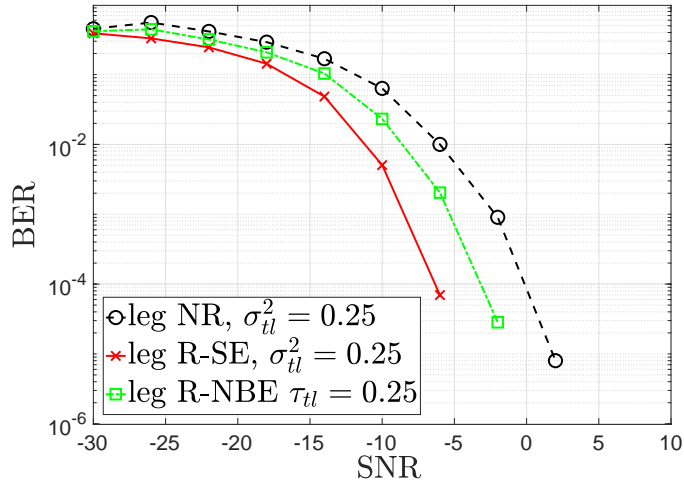}
\caption{BER at legitimate UEs (leg) vs. transmit SNR (in dB) with non-robust (NR), robust with stochastic errors (R-SE) and robust with norm-bounded errors (R-NBE) designs ($K_T=8$, $K_R=16$, $K_E=4$, eavesdroppers experience NBE with $\tau_{te}=0.09$).}
\label{fig:ber_diff_systems}
\end{figure}

\subsubsection{Effect of CSI errors}
Fig.~\ref{fig:ber_diff_systems} shows the BER as a function of the transmit SNR for robust and non-robust designs at legitimate UEs. We observe the interest of designing a system robust to CSI errors to improve the reliability of MCC: up to 6~dB gain can be achieved at low BER when assuming stochastic errors. As expected, the performance in presence of norm-bounded errors is worse than with stochastic errors. This is due to the higher noise uncertainty: noise is drawn in a sphere of known radius but there is no further statistical knowledge about it. Nevertheless, NBE-based robust design achieves a 3~dB gain at low BER.       
\begin{figure}[t]
\centering
\includegraphics[width=0.6\textwidth]{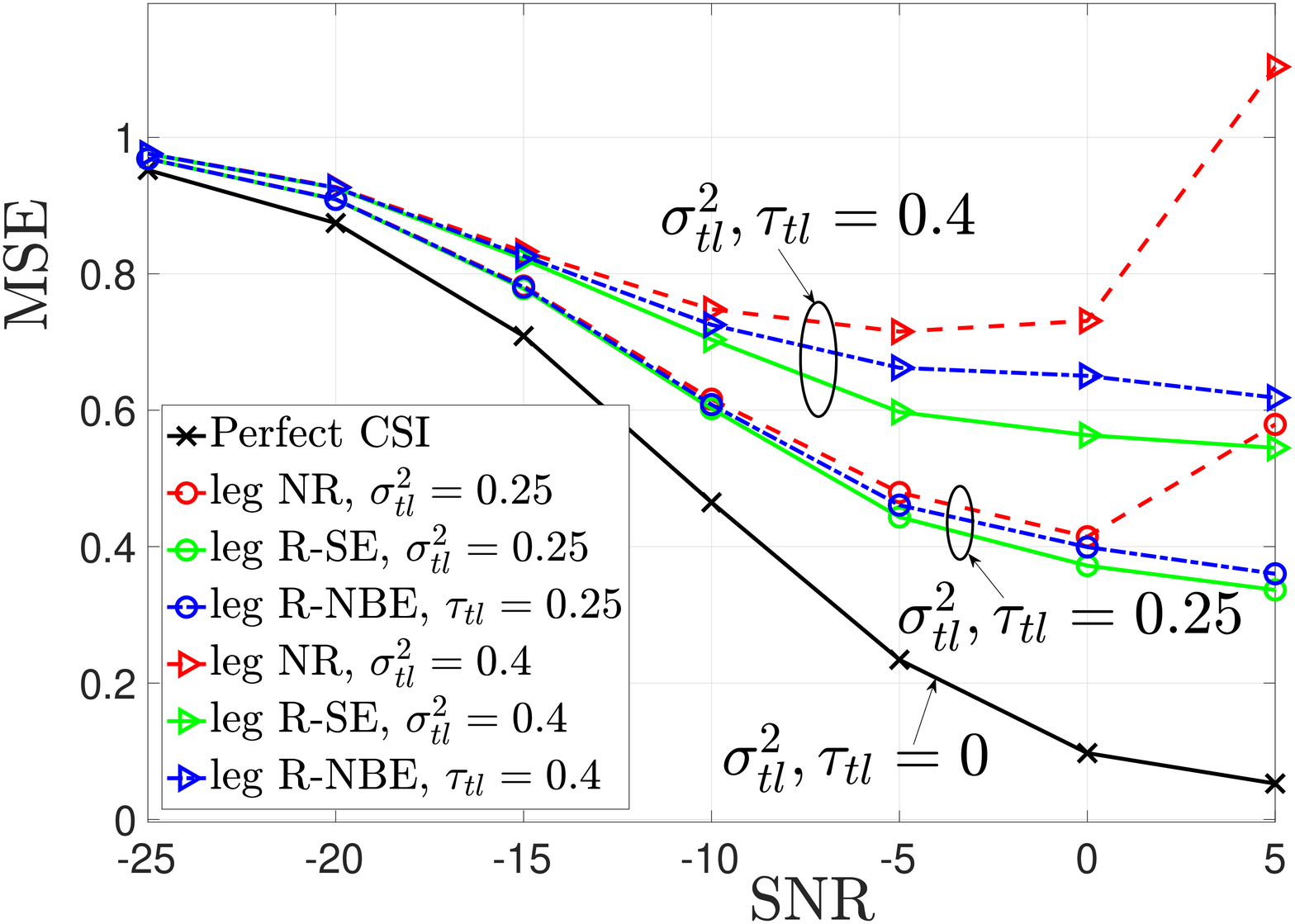}
\caption{MSE at legitimate UEs (leg) vs. transmit SNR (in dB) with perfect CSI, non-robust (NR) and robust (R) designs with stochastic errors (SE) or norm-bounded errors (NBE) ($P_T=20$~dBm, eavesdroppers experience NBE with $\tau_{te}=0.09$).}
\label{fig:error_variance_effect}
\end{figure}
Fig.~\ref{fig:error_variance_effect} shows the effect of different channel error variances on the MSE at legitimate UEs, which is our objective in the proposed minimization problems. Obviously the scenario with perfect CSI performs the best. Then, as expected, the higher the channel errors the lower the performance. The knowledge of the probability distribution function of noise (SE) is a clear advantage over sole knowledge of an upper bound (NBE) for a robust design. 

%
\subsubsection{Security Gap}
\begin{figure}[t]
\centering
\includegraphics[width=0.9\textwidth]{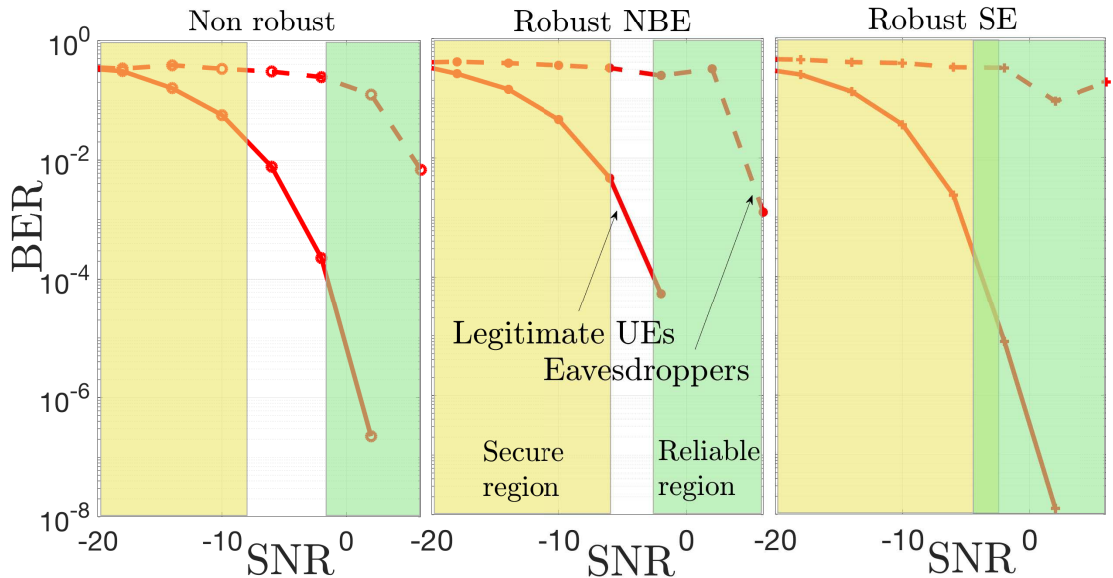}
\caption{BER vs. transmit SNR (dB) and BER Security gap for non-robust, robust with norm-bounded errors (NBE) and robust with stochastic errors (SE) (target BER of 0.3 for eavesdroppers, target BER of $10^{-4}$ for legitimate UEs).}
\label{fig:security_gap}
\end{figure}
The security gap is a measure of the secrecy level based on the BER performance of legitimate UEs and eavesdroppers. The security gap is defined as $S_g=SNR_{min}^L-SNR_{max}^E$ where $SNR_{min}^L$ is the minimum SNR at a legitimate UE to achieve high reliability (e.g. $10^{-4}$ in our simulations) and $SNR_{max}^E$ is the maximum SNR that guarantees a high BER at a eavesdropper (e.g. $0.3$). Below $SNR_{max}^E$ at eavesdroppers, the communication is secure, above $SNR_{min}^L$ at legitimate UEs, the communication is reliable. The gap quantifies the advantage a legitimate UE should have over eavesdroppers in order to have a secure and reliable communication. We see in Fig.~\ref{fig:security_gap} (left) that multi-user MIMO and artificial noise leads already to a small security gap ($4.5$~dB) even with a non-robust design. Robust designs allow an even smaller gap, especially with the stochastic error model ($2.5$~dB with NBE in the center and $-2.3$~dB with SE on the right of the figure).


\subsubsection{Effect of AN}
\begin{figure}
\centering
\includegraphics[width=0.6\textwidth]{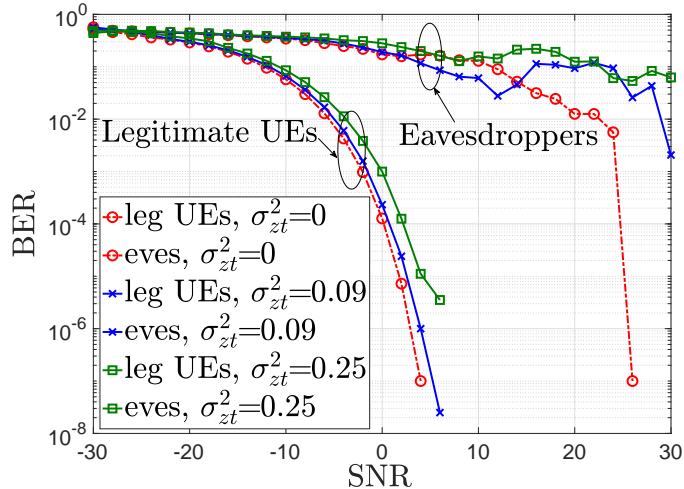}
\caption{BER vs. transmit SNR (dB) at legitimate UEs (leg) and eavesdroppers (eves) for varying AN variance for robust norm-bounded errors (R-NBE) design ($\tau_{tl}=0.04$, $\tau_{te}=0.09$).}
\label{fig:an_var_effect}
\end{figure}

Fig.~\ref{fig:an_var_effect} shows the effect of the AN variance on the BER of legitimate UEs and eavesdroppers for the R-NBE design. 
It is first observed that the BER of the legitimate UEs is significantly lower than the eavesdroppers BER, hence guaranteeing a reliable communication. The addition of AN lowers a bit the performance of legitimate UEs as some power is dedicated to it. On the contrary for eavesdroppers, there is a significant gap between the system performance with and without AN. For the specific case of multicast MCC, this confirms the interest of AN for secure communications shown in the literature in other contexts.  
\begin{figure}[t]
\centering
\includegraphics[width=0.6\textwidth]{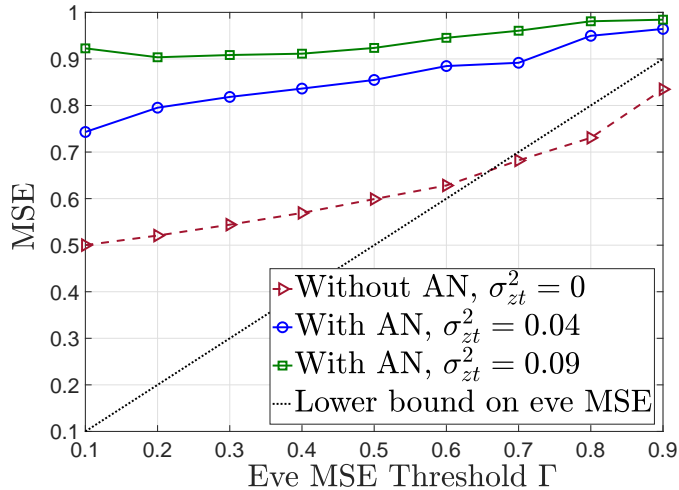}
\caption{MSE at eavesdropper vs eavesdropper MSE threshold $\Gamma$ for varying AN at $SNR=-10$~dB for robust norm-bounded errors (R-NBE) design.}
\label{fig:eve_mse_threshold}
\end{figure}
Fig.~\ref{fig:eve_mse_threshold} shows the MSE at eavesdroppers as a function of the threshold $\Gamma$ considered in our optimization problems. It is observed that the system designed without the consideration of AN is not always able to satisfy the requirement, whereas all the thresholds are readily satisfied by the AN-aided system design. Further, with the increase in the variance of AN, the system is more likely to achieve higher thresholds and inherently provides enhanced security against eavesdroppers. 



\subsection{System Level Simulations}
\label{sec:sys_level_simulations}
System level simulations allow us to account for co-channel interference and random locations of the users. We consider 100 BSs, distributed over an area of $10$~km$^2$, drawn according to a Poisson process. The synchronization area is made of $20$ BSs (see Fig.~\ref{fig:net_mod}). The path-loss between BSs and users is calculated as per Okumura-Hata model using a carrier frequency of $700$~MHz as given in \cite{hata_pl} (a typical frequency for MCC).  We assume $N_T=16$, $N_R=8$, $N_E=4$, $N_s=2$, $P_T=46$~dBm. The system considered is with robust-NBE at both legitimate and eavesdroppers with $\tau_{tl}=0.04$ and $\tau_{te}=0.09$ respectively. AN at $t$-th BS is set to $\sigma_{zt}^2=0.04$. For a simulation, a team leader is uniformly drawn in the synchronization area and then $9$ team members are selected within a distance of $500$~m. Two eavesdroppers are randomly drawn within the same distance around the team leader. All the simulations are executed over $10^6$ data streams. Simulations are performed for 100 groups. 
\begin{figure}[t]
\centering
\begin{subfigure}{0.48\textwidth}
\centering
\includegraphics[width=\textwidth]{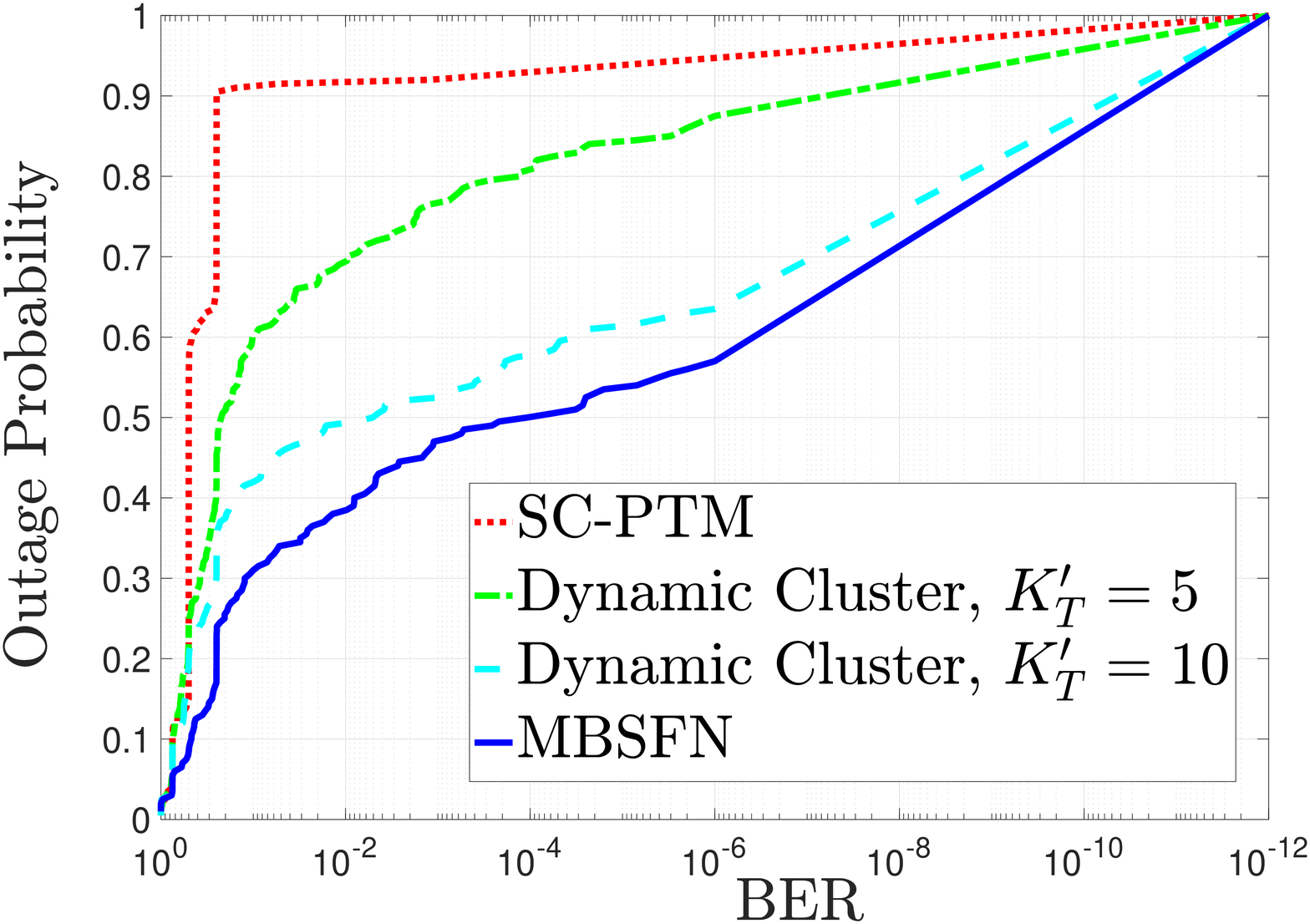}
\caption{BER CDF for legitimate UEs.}
\label{fig:system_sim_leg}
\end{subfigure}
\hfill
\begin{subfigure}{0.48\textwidth}
\centering
\includegraphics[width=\textwidth]{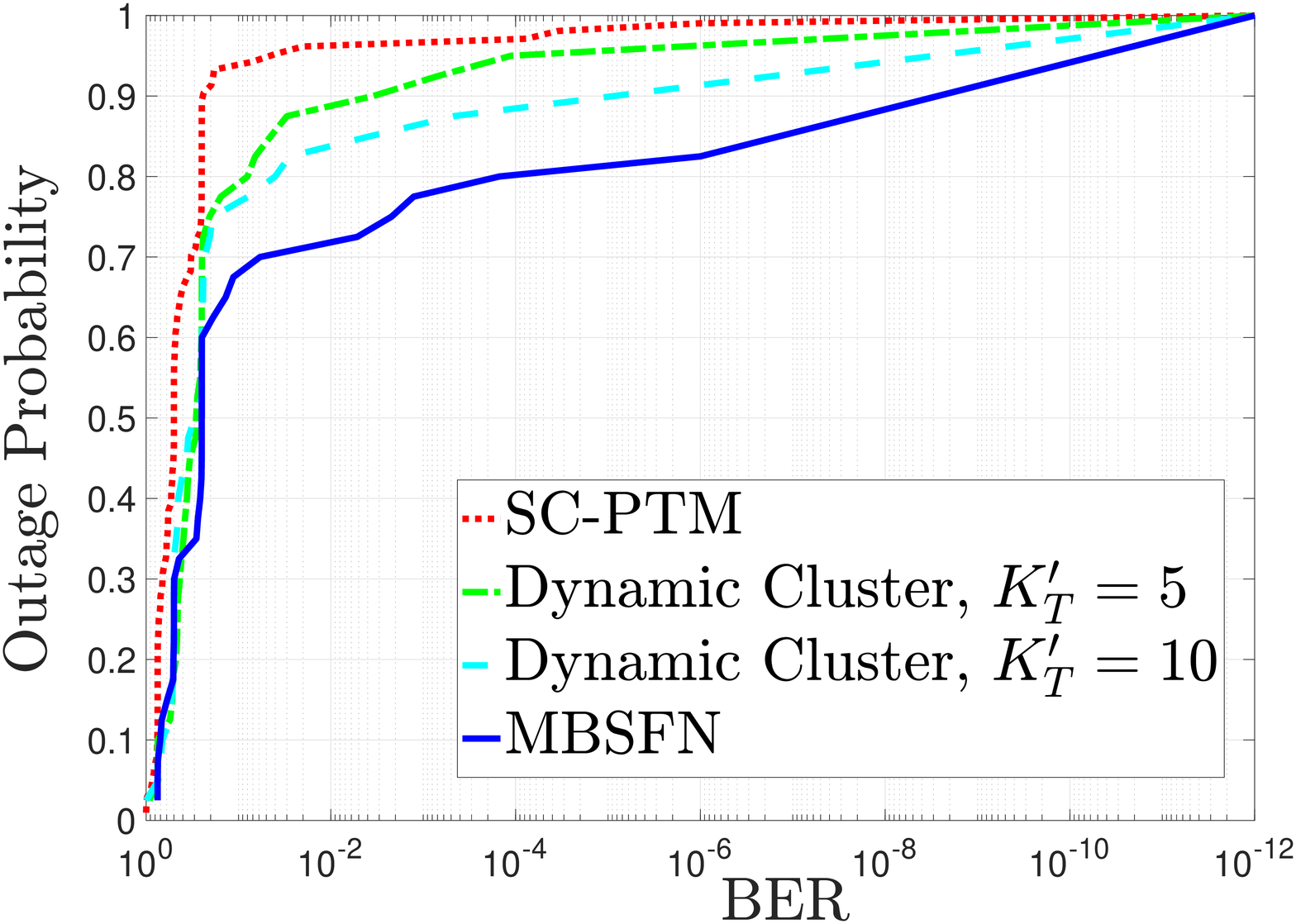}
\caption{BER CDF for eavesdroppers.}
\label{fig:system_sim_eve}
\end{subfigure}
\caption{BER CDF with MBSFN, SC-PTM and dynamic clustering.}
\label{fig:system_level_sim}
\end{figure}
Fig.~\ref{fig:system_level_sim} shows the BER CDF of legitimate UEs (left) and eavesdroppers (right) for different clustering approaches: MBSFN (the whole synchronization area serves the legitimate UEs), SC-PTM (only cells covering UEs multicast information without cooperation) and dynamic greedy clustering (Algorithm~\ref{alg:greedy}, where $K'_T$ controls the minimum number of BSs involved in the cluster). As expected MBSFN provides the best performance, SC-PTM the worst and dynamic clustering offers a tradeoff\footnote{SC-PTM provides however the best performance in terms of system capacity as studied in \cite{alaa_dynamic_clus_mcc}.}. This is true for both legitimate and eavesdroppers but the gap between the two is much higher with MBSFN compared to SC-PTM. MBSFN should thus be preferred for secure and highly reliable MCC. However, a drawback of MBSFN is that it consumes radio resources in every BS of the synchronization area and thus suffers from low capacity. If an operator wants to increase its network capacity, it should trade-off the security and reliability level against capacity by adopting a dynamic clustering scheme. Dynamic clustering with $K'_T=10$ BSs, which is half of the synchronization area represents for example here a good trade-off. 


%% file: sections/appendix_lemma1.tex
The transmit power can be expressed as follows:
\begin{eqnarray}
{P}_{t} &\triangleq&\mathbb{E}[||{\bf x}_{t}{\bf x}_{t}^{H}||] \nonumber \\
&=& \mathbb{E}[{\rm}{tr}({\bf x}_{t}{\bf x}_{t}^{H})]\nonumber \\
&=& \mathbb{E}[{\rm}{tr}(({\bf V}_{t}{\bf d} + {\bf w}_{t})({\bf V}_{t}{\bf d} + {\bf w}_{t})^{H})]\nonumber \\
&=&{\rm}{tr}({\bf V}_{t}{\bf V}_{t}^{H}+{\bf W}_{t}). 
\end{eqnarray}
The MSE ${\epsilon}_{l}$ at the legitimate user $l$ is computed as follows:
\begin{eqnarray}
{\epsilon}_{l} &\triangleq& \mathbb{E}[||{\bf d}-\widehat{\bf d}_{l}||^{2}], \nonumber  \\
&=& \mathbb {E}\bigg[||{\bf d}-(%
\sum_{t=1}^{K_T}{\bf R}_{l}(\widehat{\bf C}_{tl} + \boldsymbol{\Delta}_{tl}){\bf V}_{t}{\bf d} +%
\sum_{t=1}^{K_T}{\bf R}_{l}(\widehat{\bf C}_{tl} + \boldsymbol{\Delta}_{tl}){\bf w}_{t} +%
{\bf R}_{l}{\bf n}_{l}%
)||^{2}\bigg]\nonumber \\
&=& {\rm tr}({\bf I}) -%
{\rm tr}(\sum_{t=1}^{K_T}{\bf R}_{l}\widehat{\bf C}_{tl}{\bf V}_{t}) -%
{\rm tr}(\sum_{t=1}^{K_T}{\bf V}_{t}^{H}\widehat{\bf C}_{tl}^{H}{\bf R}_{l}^{H}) +%
{\rm tr}(\sum_{t=1}^{K_T}{\bf R}_{l}\widehat{\bf C}_{tl}{\bf V}_{t}{\bf V}_{t}^{H}\widehat{\bf C}_{tl}^{H}{\bf R}_{l}^{H})  \nonumber \\%
&+&{\rm tr}(\sum_{t=1}^{K_T}{\bf R}_{l}\widehat{\bf C}_{tl}{\bf W}_{t}\widehat{\bf C}_{tl}^{H}{\bf R}_{l}^{H}) +%
{\sigma}_{nl}^{2}{\rm tr}({\bf R}_{l}{\bf R}_{l}^{H}) +%
\mathbb{E}\big[{\rm tr}(\sum_{t=1}^{K_T}{\bf R}_{l}\boldsymbol{\Delta}_{tl}{\bf V}_{t}{\bf V}_{t}^{H}\boldsymbol{\Delta}_{tl}^{H}{\bf R}_{l}^{H})\big] \nonumber \\
&+&  \mathbb{E}\big[{\rm tr}(\sum_{t=1}^{K_T}{\bf R}_{l}\boldsymbol{\Delta}_{tl}{\bf W}_{t}\boldsymbol{\Delta}_{tl}^{H}{\bf R}_{l}^{H})\big] 
\end{eqnarray}
The last two terms can be simplified by using the trace property as given in Lemma 1 in~\cite{robust_relay2}. For any matrix ${\bf X}$ with ${\mathbb E}\{{\bf X}{\bf
X}^H\}=\sigma^2{\bf I}$, and matrices ${\bf U}$ and ${\bf V}$ of appropriate dimensions, the following equality indeed holds:
$\mathbb{E} [{\rm tr}(\mathbf{XU}\mathbf{X}^{H}\mathbf{V})] = \mathbb{E}[{\rm tr}(\mathbf{X}^{H}\mathbf{VXU})] =\sigma^{2}{\rm tr}(\mathbf{U}){\rm tr}(\mathbf{V}).\label{eq:traceproperty}$
Incorporating this result into the equation yields the result. 
Similarly, MSE at the eavesdropper $e$ can be formulated as:
\begin{eqnarray}
{\epsilon}_{e} &\triangleq&  \mathbb{E}\bigg[||{\bf d}-\overline{\bf d}_{e}||^{2}\bigg] \nonumber  \\
&=&\mathbb{E}\bigg[||{\bf d}-(\sum_{t=1}^{K_T}{\bf E}_{e}(\widehat{\bf G}_{te} + \boldsymbol{\Delta}_{te}){\bf V}_{t}{\bf d} + \sum_{t=1}^{K_T}{\bf E}_{e}(\widehat{\bf G}_{te} + \boldsymbol{\Delta}_{te}){\bf w}_{t} + {\bf E}_{e}{\bf n}_{e})||^{2}\bigg] \nonumber \\
&=& {\rm tr}({\bf I}) -%
{\rm tr}(\sum_{t=1}^{K_T}{\bf E}_{e}\widehat{\bf G}_{te}{\bf V}_{t}) -%
{\rm tr}(\sum_{t=1}^{K_T}{\bf V}_{t}^{H}\widehat{\bf G}_{te}^{H}{\bf E}_{e}^{H}) +%
{\rm tr}(\sum_{t=1}^{K_T}{\bf E}_{e}\widehat{\bf G}_{te}{\bf V}_{t}{\bf V}_{t}^{H}\widehat{\bf G}_{te}^{H}{\bf E}_{e}^{H}) \nonumber \\%
&+& {\rm tr}(\sum_{t=1}^{K_T}{\bf E}_{e}\widehat{\bf G}_{te}{\bf W}_{t}\widehat{\bf G}_{te}^{H}{\bf E}_{e}^{H}) +%
{\sigma}_{ne}^{2}{\rm tr}({\bf E}_{e}{\bf E}_{e}^{H}) +%
\mathbb{E}\big[{\rm tr}(\sum_{t=1}^{K_T}{\bf E}_{e}\boldsymbol{\Delta}_{te}{\bf V}_{t}{\bf V}_{t}^{H}\boldsymbol{\Delta}_{te}^{H}{\bf E}_{e}^{H})\big] \nonumber \\
&+&  \mathbb{E}\big[{\rm tr}(\sum_{t=1}^{K_T}{\bf E}_{e}\boldsymbol{\Delta}_{te}{\bf W}_{t}\boldsymbol{\Delta}_{te}^{H}{\bf E}_{e}^{H})\big] 
\end{eqnarray}
Again, the application of the trace property provides the result.

%% file: sections/appendix_alternativeW.tex
We use here two binary slack variables ${\chi}_{l}$ and ${\chi}_{e}$ in order to consider at once different problems introduced in the paper. ${\chi}_{l} = 1$ corresponds to a robust solution for legitimate users and ${\chi}_{l} = 0$ to a non-robust design. ${\chi}_{e} = 1$ corresponds to a perfect eavesdroppers CSI at the transmitter, otherwise ${\chi}_{e} = 0$. The generalized MSE equations are now reformulated as:
\begin{eqnarray}
{\epsilon}_{l} &=& {\rm tr}({\bf I}) -%
{\rm tr}(\sum_{t=1}^{K_T}{\bf R}_{l}\widehat{\bf C}_{tl}{\bf V}_{t}) -%
{\rm tr}(\sum_{t=1}^{K_T}{\bf V}_{t}^{H}\widehat{\bf C}_{tl}^{H}{\bf R}_{l}^{H}) +%
{\rm tr}(\sum_{t=1}^{K_T}{\bf R}_{l}\widehat{\bf C}_{tl}{\bf V}_{t}{\bf V}_{t}^{H}\widehat{\bf C}_{tl}^{H}{\bf R}_{l}^{H}) \nonumber \\%
&+& {\rm tr}(\sum_{t=1}^{K_T}\sigma_{zt}^2{\bf R}_{l}\widehat{\bf C}_{tl}{\bf W}_{t}{\bf W}_{t}^{H}\widehat{\bf C}_{tl}^{H}{\bf R}_{l}^{H}) +%
{\sigma}_{nl}^{2}{\rm tr}({\bf R}_{l}{\bf R}_{l}^{H}) +%
{\chi}_{l}\sum_{t=1}^{K_T}{\sigma}_{tl}^{2}{\rm tr}({\bf R}_{l}{\bf R}_{l}^{H}){\rm tr}({\bf V}_{t}{\bf V}_{t}^{H}) \nonumber \\%
&+& {\chi}_{l}\sum_{t=1}^{K_T}{\sigma}_{tl}^{2}{\sigma}_{zt}^{2}{\rm tr}({\bf R}_{l}{\bf R}_{l}^{H}){\rm tr}({\bf W}_{t}{\bf W}_{t}^{H}) \label{eq:mse_leg_gen}
\end{eqnarray}
The $e$-th eavesdroppers MSE is simplified as:
\begin{eqnarray}
{\epsilon}_{e} &=& {\rm tr}({\bf I}) -%
{\rm tr}(\sum_{t=1}^{K_T}{\bf E}_{e}\widehat{\bf G}_{te}{\bf V}_{t}) -%
{\rm tr}(\sum_{t=1}^{K_T}{\bf V}_{t}^{H}\widehat{\bf G}_{te}^{H}{\bf E}_{e}^{H}) +%
{\rm tr}(\sum_{t=1}^{K_T}{\bf E}_{e}\widehat{\bf G}_{te}{\bf V}_{t}{\bf V}_{t}^{H}\widehat{\bf G}_{te}^{H}{\bf E}_{e}^{H}) \nonumber \\%
&+& {\rm tr}(\sum_{t=1}^{K_T}{\sigma}_{zt}^{2}{\bf E}_{e}\widehat{\bf G}_{te}{\bf W}_{t}{\bf W}_{t}^{H}\widehat{\bf G}_{te}^{H}{\bf E}_{e}^{H}) +%
{\sigma}_{ne}^{2}{\rm tr}({\bf E}_{e}{\bf E}_{e}^{H}) +%
{\chi}_{e}\sum_{t=1}^{K_T}{\sigma}_{te}^{2}{\rm tr}({\bf E}_{e}{\bf E}_{e}^{H}){\rm tr}({\bf V}_{t}{\bf V}_{t}^{H}) \nonumber \\%
&+& {\chi}_{e}\sum_{t=1}^{K_T}{\sigma}_{te}^{2}{\sigma}_{zt}^{2}{\rm tr}({\bf E}_{e}{\bf E}_{e}^{H}){\rm tr}({\bf W}_{t}{\bf W}_{t}^{H}) \label{eq:mse_eve_gen}
\end{eqnarray}
We solve the optimization problem~\eqref{P1} by writing the Lagrangian ${\rm L}$:
\begin{eqnarray}
\label{eq:lagrange}
{\rm L}({\bf V}_{t}, {\bf W}_{t}, {\bf R}_{l}, {\lambda}_{e}, {\lambda}_{t}^{'})\hspace{-0.3cm} &=&\hspace{-0.3cm} \sum_{l=1}^{K_R}{\epsilon}_{l} - \sum_{e=1}^{K_E}{\lambda}_{e}({\epsilon}_{e} - \Gamma) + \sum_{t=1}^{K_T}{\lambda}_{t}^{'}({\rm tr}({\bf V}_{t}{\bf V}_{t}^{H} + \sigma_{zt}^{2}{\bf W}_{t}{\bf W}_{t}^{H}) - P_T) 
\end{eqnarray}
The desired transceivers are obtained by minimizing the Lagrangian with respect to each optimization variable while considering that the other variables as fixed. Hence, the precoder ${\bf V}_{t}$ is derived by taking the zero gradient of ${\rm L}$ with respect to ${\bf V}_{t}^{H}$, and is given as: 
\begin{eqnarray}
\label{eq:V_alternative}
\frac{\partial{\rm L}}{\partial{\bf V}_{t}^{H}} &=& - \sum_{l=1}^{K_R}\widehat{{\bf C}}_{tl}^{H}{\bf R}_{l}^{H} +%
\sum_{l=1}^{K_R}\widehat{{\bf C}}_{tl}^{H}{\bf R}_{l}^{H}{\bf R}_{l}\widehat{\bf C}_{tl}{\bf V}_{t} +%
{\chi}_{l}\sum_{l=1}^{K_R}{\sigma}_{tl}^{2}{\rm tr}({\bf R}_{l}{\bf R}_{l}^{H}){\bf V}_{t} +%
\sum_{e=1}^{K_E}{\lambda}_{e}\widehat{\bf G}_{te}^{H}{\bf E}_{e}^{H} \nonumber \\%
&-&\quad \sum_{e=1}^{K_E}{\lambda}_{e}\widehat{\bf G}_{te}^{H}{\bf E}_{e}^{H}{\bf E}_{e}\widehat{\bf G}_{te}{\bf V}_{t} -%
{\chi}_{e}\sum_{e=1}^{K_E}{\lambda}_{e}{\sigma}_{te}^{2}{\rm tr}({\bf E}_{e}{\bf E}_{e}^{H}){\bf V}_{t} +%
\lambda_{t}^{'}{\bf V}_{t}
\end{eqnarray}
Equatting to zero leads to:
\begin{eqnarray}
{\bf V}_{t} &=& \Big(\sum_{l=1}^{K_R}\widehat{{\bf C}}_{tl}^{H}{\bf R}_{l}^{H}{\bf R}_{l}\widehat{\bf C}_{tl} +%
{\chi}_{l}\sum_{l=1}^{K_R}{\sigma}_{tl}^{2}{\rm tr}({\bf R}_{l}{\bf R}_{l}^{H}){\bf I} - 
\sum_{e=1}^{K_E}{\lambda}_{e}\widehat{\bf G}_{te}^{H}{\bf E}_{e}^{H}{\bf E}_{e}\widehat{\bf G}_{te} \nonumber \\%
&-&\quad {\chi}_{e}\sum_{e=1}^{K_E}{\lambda}_{e}{\sigma}_{te}^{2}{\rm tr}({\bf E}_{e}{\bf E}_{e}^{H}){\bf I} +%
\lambda_{t}^{'}{\bf I}\Big)^{-1}%
{\Big(\sum_{l=1}^{K_R}\widehat{{\bf C}}_{tl}^{H}{\bf R}_{l}^{H} -%
\sum_{e=1}^{K_E}{\lambda}_{e}\widehat{\bf G}_{te}^{H}{\bf E}_{e}^{H} \Big)} \label{eq:precoder_gen}
\end{eqnarray}
Receive filter ${\bf R}_{l}$ is obtained in the same way, i.e. by differentiating the Lagrangian with respect to ${\bf R}_{l}^{H}$, while considering all other variables as fixed, and assigning it to zero:
\begin{eqnarray}
\frac{\partial{\rm L}}{\partial{\bf R}_{l}^{H}} &=& - \sum_{t=1}^{K_T}{\bf V}_{t}^{H}\widehat{{\bf C}}_{tl}^{H} +%
\sum_{t=1}^{K_T}{\bf R}_{l}\widehat{{\bf C}}_{tl}{\bf V}_{t}{\bf V}_{t}^{H}\widehat{{\bf C}}_{tl}^{H} +%
\sum_{t=1}^{K_T}\sigma_{zt}^{2}{\bf R}_{l}\widehat{{\bf C}}_{tl}{\bf W}_{t}{\bf W}_{t}^{H}\widehat{{\bf C}}_{tl}^{H} +%
{\sigma}_{nl}^{2}{\bf R}_{l}  \nonumber \\%
&+&\quad {\chi}_{l}\sum_{t=1}^{K_T}{\sigma}_{tl}^{2}{\rm tr}({\bf V}_{t}{\bf V}_{t}^{H}){\bf R}_{l} +%
{\chi}_{l}\sum_{t=1}^{K_T}{\sigma}_{tl}^{2}\sigma_{zt}^{2}{\rm tr}({\bf W}_{t}{\bf W}_{t}^{H}){\bf R}_{l} \\
{\bf R}_{l} &=& \Big(\sum_{t=1}^{K_T}{\bf V}_{t}^{H}\widehat{{\bf C}}_{tl}^{H}\Big)\Big(\sum_{t=1}^{K_T}\widehat{{\bf C}}_{tl}{\bf V}_{t}{\bf V}_{t}^{H}\widehat{{\bf C}}_{tl}^{H} +%
\sum_{t=1}^{K_T}\sigma_{zt}^{2}\widehat{{\bf C}}_{tl}{\bf W}_{t}{\bf W}_{t}^{H}\widehat{{\bf C}}_{tl}^{H} + {\sigma}_{nl}^{2}{\bf I} \nonumber \\%
&+&\quad {\chi}_{l}\sum_{t=1}^{K_T}{\sigma}_{tl}^{2}{\rm tr}({\bf V}_{t}{\bf V}_{t}^{H}){\bf I} +%
{\chi}_{l}\sum_{t=1}^{K_T}{\sigma}_{tl}^{2}{\sigma}_{zt}^{2}{\rm tr}({\bf W}_{t}{\bf W}_{t}^{H}){\bf I} \Big)^{-1} \label{eq:receiver_gen}
\end{eqnarray}
Now, we differentiate the Lagrangian with respect to ${\bf W}_{t}^{H}$ and assign it to zero:
\begin{eqnarray}
\label{eq:an_gen}
\frac{\partial{\rm L}}{\partial{\bf W}_{t}^{H}} &=& \sum_{l=1}^{K_R}\sigma_{zt}^{2}\widehat{{\bf C}}_{tl}^{H}{\bf R}_{l}^{H}{\bf R}_{l}\widehat{\bf C}_{tl}{\bf W}_{t} +%
{\chi}_{l}\sigma_{zt}^{2}\sum_{l=1}^{K_R}{\sigma}_{tl}^{2}{\rm tr}({\bf R}_{l}{\bf R}_{l}^{H}){\bf W}_{t} -%
\sigma_{zt}^{2}\sum_{e=1}^{K_E}{\lambda}_{e}\widehat{\bf G}_{te}^{H}{\bf E}_{e}^{H}{\bf E}_{e}\widehat{\bf G}_{te}{\bf W}_{t} \nonumber \\%
&-&\quad {\chi}_{e}\sigma_{zt}^{2}\sum_{e=1}^{K_E}{\lambda}_{e}{\sigma}_{te}^{2}{\rm tr}({\bf E}_{e}{\bf E}_{e}^{H}){\bf W}_{t}+%
\lambda_{t}^{'}\sigma_{zt}^{2}{\bf W}_{t} 
\end{eqnarray}
\begin{eqnarray}
0&=& \sigma_{zt}^{2}\Bigg[
 \sum_{l=1}^{K_R}\widehat{{\bf C}}_{tl}^{H}{\bf R}_{l}^{H}{\bf R}_{l}\widehat{\bf C}_{tl} + {\chi}_{l}\sum_{l=1}^{K_R}{\sigma}_{tl}^{2}{\rm tr}({\bf R}_{l}{\bf R}_{l}^{H})-\sum_{e=1}^{K_E}{\lambda}_{e}\widehat{\bf G}_{te}^{H}{\bf E}_{e}^{H}{\bf E}_{e}\widehat{\bf G}_{te}\\ \nonumber 
&&  -{\chi}_{e}\sum_{e=1}^{K_E}{\lambda}_{e}{\sigma}_{te}^{2}{\rm tr}({\bf E}_{e}{\bf E}_{e}^{H})+ \lambda_{t}^{'}{\bf I}
\Big] {\bf W}_{t}
\end{eqnarray}
This is equivalent to ${\bf A}_{t}{\bf W}_{t}=0$
where
\begin{eqnarray}
{\bf A}_{t}&=& \sum_{l=1}^{K_R}\widehat{{\bf C}}_{tl}^{H}{\bf R}_{l}^{H}{\bf R}_{l}\widehat{\bf C}_{tl} + {\chi}_{l}\sum_{l=1}^{K_R}{\sigma}_{tl}^{2}{\rm tr}({\bf R}_{l}{\bf R}_{l}^{H})-\sum_{e=1}^{K_E}{\lambda}_{e}\widehat{\bf G}_{te}^{H}{\bf E}_{e}^{H}{\bf E}_{e}\widehat{\bf G}_{te} \nonumber \\
&-&{\chi}_{e}\sum_{e=1}^{K_E}{\lambda}_{e}{\sigma}_{te}^{2}{\rm tr}({\bf E}_{e}{\bf E}_{e}^{H})+ \lambda_{t}^{'}{\bf I}
\end{eqnarray}
In the condition ${\bf A}_{t}{\bf W}_{t}=0$, ${\bf A}_{t}$ cannot be zero because otherwise effective components in the design of the precoder would be zero and the precoder matrix would be non-singular. This would invalidate the complete design. As a consequence, the AN shaping matrix ${\bf W}_t$ should be taken in the null space of ${\bf A}_{t}$ and: ${\bf W}_t = {\bf B}_{t}/\sqrt{\mathbb[\rm tr({\bf B}_{t}{\bf B}_{t}^{H})]}$ where ${\bf B}_{t} = {\bf I}-{\bf A}_{t}^{H}({\bf A}_{t}{\bf A}_{t}^{H})^{-1}{\bf A}_{t}$.
At last, differentiating the Lagrangian with respect to $\lambda_{t}^{'}$ and $\lambda_{e}$ respectively and equating to zero, we get:
\begin{eqnarray}
\frac{\partial{\rm L}}{\partial{\lambda_{t}^{'}}} &=& {\rm tr}({\bf V}_{t}{\bf V}_{t}^{H} +\sigma_{zt}^{2}{\bf W}_{t}{\bf W}_{t}^{H}) - P_{T} = 0 \label{eq:lambda_alternative}\\
\frac{\partial{\rm L}}{\partial{\lambda_{e}}} &=&
{\epsilon}_e - \Gamma = 0 \label{eq:lambda_e}
\end{eqnarray}
The values for $\lambda_{e}$, $e =\{1,2,\cdots,K_E\}$ and $\lambda_{t}^{'}$, $t =\{1,2,\cdots,K_T\}$ are jointly computed by inserting the values of ${\bf V}_t$ and ${\bf W}_t$ in \eqref{eq:lambda_alternative} and \label{eq:lambda_e} so as to satisfy the transmit power constraint at each BS and individual eavesdroppers MSE threshold.